%% file: paper_revision_heath_v2.tex
\documentclass[draftclsnofoot, 12pt, onecolumn]{IEEEtran}
\normalsize
\input{input.tex}
\usepackage{mathtools}
\usepackage{bbm}
\usepackage{amsthm}
\usepackage{cite}
\usepackage{diagbox}
\usepackage{url}

\DeclarePairedDelimiter\floor{\lfloor}{\rfloor}

\graphicspath{{figure/}}
\title{Site-specific online compressive beam codebook learning in mmWave vehicular communication}

\author{\IEEEauthorblockN{Yuyang Wang, Nitin Jonathan Myers, Nuria Gonz\'alez-Prelcic, \\
and Robert W. Heath Jr. }\thanks{Yuyang Wang, Nitin Jonathan Myers, Nuria Gonz\'alez-Prelcic, and Robert W. Heath Jr. are with the Department of Electrical and Computer Engineering, the University of Texas at Austin, Austin, TX, 78712 USA, email: $\{$yuywang, nitinjmyers, ngprelcic, rheath$\}$@utexas.edu. This research was partially supported by a gift from FutureWei through UT Situation-Aware Vehicular Engineering Systems (UT-SAVES). }}

\begin{document}
\maketitle 
\vspace{-5pt}
\begin{abstract}
Millimeter wave (mmWave) communication is one viable solution to support Gbps sensor data sharing in vehicular networks. The use of large antenna arrays at mmWave and high mobility in vehicular communication make it challenging to design fast beam alignment solutions. In this paper, we propose a novel framework that learns the channel angle-of-departure (AoD) statistics at a base station (BS) and uses this information to efficiently acquire channel measurements. Our framework integrates online learning for compressive sensing (CS) codebook learning and the optimized codebook is used for CS-based beam alignment. We formulate a CS matrix optimization problem based on the AoD statistics available at the BS. Furthermore, based on the CS channel measurements, we develop techniques to update and learn such channel AoD statistics at the BS. We use the upper confidence bound (UCB) algorithm to learn the AoD statistics and the CS matrix. Numerical results show that the CS matrix in the proposed framework provides faster beam alignment than standard CS matrix designs. Simulation results indicate that the proposed beam training technique can reduce overhead by 80$\%$ compared to exhaustive beam search, and 70$\%$ compared to standard CS solutions that do not exploit any AoD statistics. \end{abstract}
\section{Introduction}\label{sec:introduction}
With the availability of a large amount of bandwidth, mmWave communication can support massive sensor data sharing in vehicular networks \cite{choi2016millimeter, heath2016overview, rappaport2013millimeter}. Due to the large antenna arrays deployed at mmWave and the high mobility in vehicular settings, fast and efficient beam alignment strategies have to be designed \cite{rappaport2014millimeter}. For example, exhaustive search-based beam alignment where a radio scans through all the beams from a pre-defined codebook can result in substantial training overhead. The exhaustive search-based solution may not be suitable for mmWave vehicular communication due to a large number of candidates in typical beam codebooks. The mobile nature of terminals in a vehicular setting can impose a strict timing constraint for the beam alignment problem. As a result, exhaustive search-based beam alignment may not meet requirements like low latency and high reliability expected in mmWave vehicular networks \cite{festag2015standards}, \cite{shah20185g}. 
\par Fast beam alignment in mmWave communication can be achieved through compressed sensing (CS) or side information-aided beam search. CS-based beam alignment solutions exploit sparsity of mmWave channels and require less training overhead than exhaustive search \cite{marzi2016compressive}, \cite{ali2017millimeter}. The performance of CS-based solutions depends on the choice of the CS matrix used to obtain channel measurements. Side information, \emph{ubiquitous} in vehicular communication, can be exploited for efficient mmWave channel estimation or beam alignment \cite{santa2009sharing, papadimitratos2009vehicular,va2017inverse, wang:2019access, ali2017millimeter, gonzalez2017millimeter, myers2020deep}. Different types of side information can be obtained from sensors mounted on vehicles, including radar, LiDAR and GPS \cite{ali2019passive, ali2019millimeter, gonzalez2016radar, klautau2019lidar, garcia2016location}. Side information can be shared among vehicles by different wireless protocols such as dedicated short-range communication (DSRC), LTE or mmWave communication \cite{kenney2011dedicated, chen2017vehicle, choi2016millimeter}. %It was shown in .... that side information can help reduce mmWave vehicular beam search overhead. (clubbed refs in a similar sentence 
Situational awareness of vehicles can also be leveraged to extract environment information and further narrow down the beam search space and reduce beam training overhead \cite{wang2018mmwave, wang:2019access, klautau20185g, va2017inverse, va2019online}. %The information about the vehicular environment makes it possible to estimate non-line-of-sight (NLOS) components of the channel. - may not be needed
One limitation of the side information-aided beam selection solutions is that they rely on fully-connected vehicles to share sensor data and enable sensor fusion. Second, none of the approaches leverage the spatial structures associated with vehicular channels for beam selection. \iffalse 
Robustness to misalignment remains a challenge with large antenna array, limited sensor precision, and low percentage of connected vehicles in the cell coverage range.
\fi
%\par  Prior work has used random IID phase shift-based matrices for CS-based beam alignment . CS techniques that use the random phase shift design, however, may result in a high complexity when large antenna arrays are applied. Structured CS algorithms can be promising candidates as they can perform sparse recovery with lower computational complexity.
 %\par Machine learning can be a promising candidate for channel estimation \cite{ye2017power, o2017introduction}, and beam alignment \cite{wang:2019access, va2017inverse}. Well-equipped with cloud access and edge-computing capability, BS can learn to design the system more efficiently and make smarter decisions using the vast amount of previous data transmission records. Machine learning models were applied in \cite{wang:2019access} and \cite{wang2018mmwave} to estimate the channel and the predict optimal beam alignment based on situational awareness. Statistical learning was leveraged in \cite{va2017inverse} to learn the multi-path fingerprints in mmWave vehicular communication given receiver location. % In addition, deep learning was shown to able to solve various problems in wireless physical layer  \cite{gruber2017deep}, \cite{wang2017deep},  \cite{o2017introduction}. 

 \par  
 Vehicles channels are temporally correlated, which makes online learning a potential candidate for link configuration in vehicular settings \cite{mairal2010online, kivinen2004online, booth2019multi}. Using online learning, a radio can dynamically adapt to the new channel statistics. In \cite{sim2018online}, beam selection was performed using vehicle arrival directions.  In \cite{va2019online} and \cite{hashemi2018efficient}, receiver location was used for online beam selection in mmWave vehicular communication. Such online learning-based beam alignment solutions, however, do not exploit the sparsity of mmWave channels. Algorithms that integrate online learning-based solutions with sparsity-aware solutions may further reduce the training overhead.
% \iffalse Reinforcement learning, as another branch of online learning, was also widely investigated in wireless protocol stack design \cite{yau2012reinforcement, liu2006rl, jiang2016machine}.
 %\fi
\par 
%Compressive sensing (CS)-based beam alignment solutions exploit sparsity of mmWave channels and have a lesser training overhead than exhaustive search \cite{alkhateeb2015compressed}. The channel measurements in CS are obtained by projecting the channel on a lower-dimensional subspace with a CS matrix \cite{baraniuk2007compressive, donoho2006compressed}. The projections are subsequently used to recover the channel using optimization techniques. The performance of channel recovery depend on the choice of the CS matrix used to obtain these projections. Prior work has used random IID phase shift-based matrices for CS-based beam alignment \cite{marzi2016compressive}, \cite{rodriguez2018frequency}. CS techniques that use the random phase shift design, however, may result in a high complexity when large antenna arrays are applied. Structured CS algorithms can be promising candidates as they can perform sparse recovery with lower computational complexity.
\par In this paper, we assume a vehicular communication setting where a BS serves vehicles in its coverage area. We propose an online beam alignment framework that leverages channel sparsity and the angle domain statistics of the channel.  Using our framework, the BS learns the AoD distribution of the paths corresponding to the vehicles in its coverage area. The AoD distribution learned at the BS is further used to optimize the CS matrix. It is important to note that the proposed approach only uses CS-based channel measurements acquired at mmWave, which are used for beam alignment. As a result, our framework does not require fully-connected vehicles, frequent sensor or out-of-band information sharing \cite{va2019online,hashemi2018efficient, ali2017millimeter}. We assume that AoD distribution learning and CS-based computations are implemented at the centralized infrastructure, i.e., the BS. The main contributions of our work can be summarized as follows. \begin{itemize}
\item We propose a novel solution for mmWave beam alignment leveraging an online learning-based CS matrix design. We develop a 2D-convolutional CS (2D-CCS) technique that optimize the CS matrix based on the AoD distribution learned at the BS. According to the CS matrix design, the BS applies beam training vectors for the users to acquire compressive channel measurements. Subsequently, the BS updates the AoD statistics based on the channel feedback received from the users. The optimized CS matrix can be obtained when the estimation of AoD distribution converges.
\item We show that online CS matrix learning using our framework is analogous to a MAB problem \cite{gittins2011multi}.
\iffalse Specifically, the different beam directions can be treated as the \emph{machines} to play in MAB. \fi We use the UCB algorithm for online CS matrix learning and investigate different exploration-exploitation methods to estimate the AoD distribution \cite{garivier2011upper}. \iffalse The estimated angular distribution can be fairly \emph{biased} in the beginning with only few samplings. Directly using the biased angular distribution to design the sensing matrix will lead to error propagation over time and convergence to an erroneous angular distribution estimation. 
\fi Furthermore, we propose solutions to {calculate} the CS matrix using a variation of the estimated AoD distribution with a reduced number of measurements. We apply different exploration terms to UCB based on the levels of \emph{confidence} in the AoD distribution learning. 
\item We validate our proposed solution with comprehensive simulations. We demonstrate superior performance of the proposed solution over \emph{exhaustive beam search} and \emph{conventional CS} using the exact AoD prior. We show that the proposed online learning-based AoD distribution estimation technique converges to the ground truth using a \emph{statistical distance} metric. Last, we evaluate the effectiveness of the proposed exploration-exploitation methods. 
 \end{itemize}
We would like to highlight that our online learning-based solution is different from the common adaptive CS setting in which the rows of a CS matrix are modified during signal acquisition \cite{ji2008bayesian}. Such an approach requires continuous feedback and may result in a  substantial training overhead. In our framework, the CS matrix at the BS remains constant during the beam alignment process of any receiver, and is updated across different channel instances. 

The rest of the paper is organized as follows. The problem is motivated in Section \ref{sec:motivation}, which demonstrates
a specific structure in the AoD distribution at the BS that can be leveraged for data-driven CS in mmWave vehicular communication. The system model is explained in Section \ref{sec:datacollection}. The channel model is provided in Section \ref{sec:channelmodel}. A new CS solution that is well-suited to the AoD prior is proposed in Section \ref{sec:CS}. The online CS matrix learning problem is introduced in Section \ref{sec:onlinesensing}. The application of UCB for the online sensing matrix learning is shown in Section \ref{sec:UCBbasics} and Section \ref{sec:sensingUCB}. The exploration-exploitation tradeoff is explained in Section \ref{sec:eetradeoff}. Comprehensive numerical results are demonstrated in Section \ref{sec:numerical}. The final conclusions are drawn in Section \ref{sec:conclusion}. 
\par \textbf{Notation}$:$ $\mathbf{A}$ is a matrix, $\mathbf{a}$ is a column vector and $a, A$ denote scalars. Using this notation $\mathbf{A}^T,\overline{\mathbf{A}}$ and $\mathbf{A}^{\ast} $ represent the transpose, conjugate and conjugate transpose of $\mathbf{A}$. \iffalse We use $\mathrm{diag}\left(\mathbf{a}\right)$ to denote a diagonal matrix with entries  of $\mathbf{a}$ on its diagonal. \fi The scalar ${\mathrm{a}}_m$ or $\mathbf{a}[m]$ denotes the $m^{\mathrm{th}}$ element of $\mathbf{a}$. The $\ell_2$ norm of $\mathbf{a}$ is denoted by $\Vert \mathbf{a} \Vert_2$. \iffalse   The $k^{\mathrm{th}}$ row and the $\ell^{\mathrm{th}}$ column of $\mathbf{A}$ are denoted by $\mathbf{A}(k,:)$ and $\mathbf{A}(:,\ell)$. \fi The scalar $\mathbf{A}\left(k,\ell\right)$ denotes the entry of $\mathbf{A}$ in the $k^{\mathrm{th}}$ row and ${\ell}^{\mathrm{th}}$ column. The matrix $|\mathbf{A}|$ contains the element-wise magnitude of $\mathbf{A}$, i.e., $|\mathbf{A}|_{k, \ell}=|\mathbf{A}_{k,\ell}|$. The $\ell_1$ norm of $\mathbf{A}$ is denoted by $\Vert \mathbf{A} \Vert_1$. The Frobenius norm is denoted by $\Vert\mathbf{A}\Vert_\mathrm{F}$. The inner product of two matrices $\mathbf{A}$ and $\mathbf{B}$ is defined as $\langle \mathbf{A},\mathbf{B}\rangle =\sum_{k,\ell}\mathbf{A}\left(k,\ell \right)\overline{\mathbf{B}}\left(k,\ell\right)$. \iffalse 
We use $\mathbf{1}$ to denote an all-ones matrix and $\mathbf{I}$ to denote the identity matrix. 
\fi
The symbols $\odot$ and $\circledast$ are used for the Hadamard product and 2D circular convolution \cite{imageprocess}. The set $[N]$ denotes the set of integers $\left\{ 0,1,2,...,N-1\right\}$. The matrix $\mathbf{U}_N \in \mathbb{C}^{N \times N}$ denotes the unitary Discrete Fourier Transform (DFT) matrix. Denote $\omega_N = e^{-2\mathsf{j}\pi/N}$, we define the DFT matrix as $\mathbf{U}_N(j, k) = \frac{\omega^{jk}_N}{\sqrt{N}}, ~j, k\in[N]$. Cross product of two sets is denoted as $\times$. 
 \section{Motivation and dataset establishment}\label{sec:motivation}
In this section, we first explain the motivation of the work. We then describe our ray tracing simulation setup developed to collect vehicular channel data. Finally, we describe how the channels are computed from the ray tracing data to generate the channel dataset. %explain the post-processing of propagation path output from ray tracing to calculate the channel and establish the dataset. \subsection{Motivation}\label{sec:motivation}

Channel statistics in vehicular communication follows regular patterns that are dependent on the site-specific environments  \cite{hassan2002new, seidel1994site, lim2017map}. We consider a scenario where a BS is deployed at road side to provide data services to passing-by vehicles that are equipped with mmWave communication devices. In most cellular scenarios, users can be located anywhere in the \emph{whole} coverage area. For vehicular communication, however, vehicles travel on predetermined lanes that occupy only a \emph{small portion} of the entire coverage area. In this case, the distribution of vehicle locations follows a statistical pattern that is related to the road layouts. Such statistical pattern of vehicle location distribution provides some informative and useful prior on the AoD corresponding to BS-vehicle links. \iffalse The AoD prior is structured for both the line-of-sight (LOS) and non line-of-sight (NLOS) links. \fi
\iffalse  Due to a deterministic mapping between the AoD corresponding to a LoS propagation path and the location of a vehicle, the AoD distribution defined by LoS paths can be very informative. The NLoS propagation paths in a vehicular environment can happen due to reflections at buildings, lamp-posts, roads, and even neighboring vehicles. As the locations of the reflectors are either fixed or have a structured distribution (in case of reflections at vehicles), the AoD prior associated with the NLoS paths is also structured based on the geometry of the street canyon. 
\fi
\par The availability of an informative AoD prior in mmWave vehicular communication settings can be helpful for efficient beam alignment or channel estimation algorithms design. For example, weighted sparse recovery techniques that exploit such a prior can be used instead of standard CS algorithms for mmWave link configuration \cite{ali2017millimeter}. Furthermore, the beam training vectors used to acquire channel measurements can be optimized to increase the probability of successful alignment. Determining the AoD prior in vehicular communication scenarios, however, can be challenging. A ray tracing-based approach to compute the AoD prior can be prohibitive in practical settings, as it may require accurate models for the surfaces of all the reflectors. The framework proposed in this paper learns the AoD prior online using compressed channel measurements, and can potentially adapt to new environments.%The AoD distribution of line-of-sight propagation paths is also structured in vehi
\subsection{Simulation setup}\label{sec:datacollection}
\begin{figure}[!h]
\centering
\includegraphics[width = 3.0in]{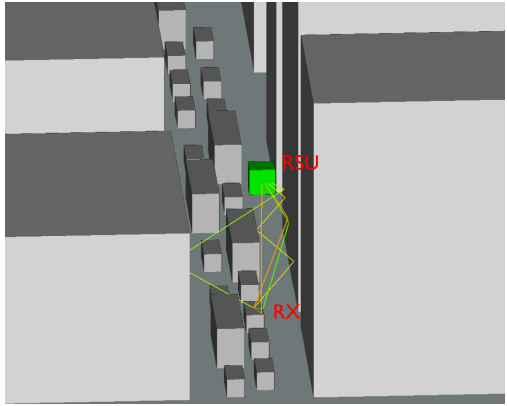}
\caption{Illustration of the ray tracing setup in which cars and trucks are randomly dropped in the two lanes of the urban canyon. Receivers are mounted on the top center of the low vehicle. The BS is mounted on a street-side lamp-post. The figure illustrates the \emph{top five} strongest paths of the channel for a certain receiver. Our channel model includes the effect of multiple reflections that occur at the buildings and the vehicles. }\label{fig:simulationsetup}
\end{figure}

We use Wireless Insite, a commercial ray tracing simulator, to establish the channel dataset for evaluation \cite{ray_tracing}. Ray tracing simulation projects rays from the BS to the physical environment and calculates the channel path information, including path gain, angle-of-arrival (AoA) and AoD between the BS and the vehicles. We consider a street layout with two straight lanes in an urban canyon. In the simulation, buildings are modeled as cuboids with a concrete exterior, located on both road sides with different dimensions. The simulation includes vehicles of different types (large vehicles like trucks and small vehicles such as sedans) randomly dropped on the two lanes. The complexity of ray tracing simulation scales exponentially with the number of surfaces. Hence, all the vehicles are modeled as cuboids with \emph{metal} exteriors for simplicity. \iffalse  In each simulation, the type of vehicle (truck and car) is determined by a Bernoulli random variable with a predefined probability. The distance between adjacent vehicles is modeled by an Erlang distribution with parameter $(k_\mathrm{erl}, \theta_\mathrm{erl})$ to form a sequence of vehicles on both lanes \cite{va2017inverse}, \cite{naumova2014model}. \fi Complete details of the ray tracing setup can be found in \cite{wang:2019access}. Our framework can be applied in \emph{any} realistic vehicular environment as it learns the underlying AoD distribution from the channel measurements. We assume that the BS is mounted on top of street-side lamp-posts, which are higher than buses and are dotted around in cities. Receivers are placed on the roof of the vehicles. A particular realization of this simulation setup is shown in  in Fig. \ref{fig:simulationsetup}. 
\subsection{Channel model}\label{sec:channelmodel}
We assume that the BS is equipped with a uniform planar array (UPA) of size $N_\mathrm x\times N_\mathrm y$. For ease of notation, we assume that the UPA has the same number of antennas along the azimuth and elevation dimensions, i.e., $N_\mathrm x =  N$ and $N_\mathrm y = N$. The derivation, however, is not subject to the square antenna constraint. We consider an analog beamforming architecture at the BS. In such an architecture, every antenna of the UPA is connected to a single radio frequency chain through a phase shifter. The use of phase shifters allows the BS to generate a variety of beams that can be used for initial access or data transmission. As the focus of this work is on the transmit beam alignment problem, we assume a single antenna receiver at the vehicles. \iffalse Our framework can also be applied to multi-antenna receivers that use a fixed beam pattern during the transmit beam alignment phase. \fi Therefore, under the multiple-input single output (MISO) system assumption, the antenna domain channel can be represented as a vector $\rh \in \mathbb{C}^{N^2 \times 1}$. The channel vector $\rh$ can be reshaped into a matrix $\rH\in\mathbb{C}^{N\times N}$. The rows and the columns of $\rH$ correspond to the elevation and the azimuth dimensions of the UPA. Note that the $(r,c)^{\mathrm{th}}$ entry of $\mathbf{H}$ represents the channel coefficient between the $(r,c)^{\mathrm{th}}$ antenna at the transmitter and the receiver antenna. The matrix representation of the MISO channel allows us to better explain the ideas underlying our beam training design. We consider both LOS and NLOS channels in our simulations.
\par We assume that the channel can be modeled using $L_\mathrm p$ paths in the propagation environment. We denote the channel gain of the $\ell$-th path as $\alpha_\ell$, the phase as $\beta_\ell$, the azimuth AoD as $\phi_\ell$, and the elevation AoD as $\theta_\ell$. We assume a half-wavelength spaced UPA at the BS to define the Vandermonde vector $\mathbf{a}_N(\Delta)$ as 
\begin{align}
\mathbf{a}_N(\Delta) =  [1, ~e^{\mathsf{j}\pi\Delta}, ~e^{2\mathsf{j}\pi\Delta}, ~\cdots, ~e^{(N - 1)\mathsf{j}\pi\Delta}]. 
\end{align}
Assuming the channel is narrowband, the MISO channel between the BS and the receiver can be expressed as 
\begin{align}
\label{eq:channel_expansion}
\rH= \sum_{\ell = 1}^{L_\mathrm{p}} \alpha_\ell e^{\mathsf{j}\beta_\ell}\mathbf{a}_N(\cos\theta_\ell)\mathbf{a}_N{(\sin\theta_\ell\cos\phi_\ell)}^T. 
\end{align}
It should be noted that we can extend our framework to the wideband case using the frame structure in \cite{myers2019falp}. The techniques proposed in this paper estimate the $N\times N$ channel matrix in \eqref{eq:channel_expansion}. This approximation is used to update the AoD distribution, which is further used for efficient beam alignment.  
\par \iffalse We use $M$ to denote the number of compressed channel measurements of the channel $\rH$ obtained by the receiver. \fi The BS applies different phase shift matrices to its antenna array in $M\ll N^2$ successive training slots for channel estimation. The receiver measures the projections of the channel on each of the phase shift matrices used by the BS. Finally, the receiver feedbacks the vector of $M$ channel measurements to the BS for beam alignment in the control channel. In the $m$-th measurement slot, $1\leq m \leq M$, the phase shift matrix is represented as $\rP[m] \in \mathbb{C}^{N\times N}$. The matrix $\rP[m]$ has a unit norm and satisfies the constant amplitude constraint. Defining $\mathbf{1}$ as an all-one matrix of size $N\times N$, we have $\left|\rP\right| = \mathbf{1}/N$. As a result, $\Vert \rP \Vert_{\mathrm F}=1$. Denoting the additive white Gaussian noise as $v[m]\sim\mathcal{N}_\mathrm c(0, \sigma^2)$ , the received signal can be written as  
\begin{align}
\label{eq:sysmodel}
y[m] = \langle \rH, \rP[m]\rangle + v[m].
\end{align}
The number of channel measurements required to estimate the $N\times N$ matrix $\rH$ can be reduced by using techniques that exploit available statistical pattern in $\rH$. 
\par The channel matrix is approximately sparse in the 2D-DFT dictionary at mmWave carrier frequencies \cite{heath2016overview} when a UPA is assumed. We define the beamspace channel $\rX\in\mathbb{C}^{N\times N}$ as the inverse 2D-DFT of $\rH$, i.e., 
\begin{align}
\rX = \rU_N^*\rH\rU_N^*. 
\end{align}
The beamspace contains the \emph{effective} channel coefficients seen when different 2D-DFT-based beams are used at the BS \cite{brady2013beamspace}. Thanks to the limited scattering characteristics at mmWave frequencies, the beamspace $\rX$ is approximately sparse. The support of the non-zero coefficients in $\rX$ is related to the AoD of the propagation rays in the channel. The sparsity of $\rX$ allows the use of CS-based algorithms for fast channel estimation or beam alignment. For the tractability of analysis, we assume that $\rX$ is one-sparse. \iffalse Under the one-sparse assumption on $\rX$, we use the terms ``AoD prior" and the ``prior on $\rX$" interchangeably. \fi The one-sparse prior for $\rX$ is perhaps simplistic, but allows a tractable design of a CS matrix that is well-suited to the prior on $\rX$. In particular, the prior is only updated based on the strongest channel component, i.e., the direction that is the strongest in the 2D-DFT beamspace. As a result, our approach cannot learn all the components of the channel and we leave the extension to the future work. The proposed CS matrix design, however, achieves good performance over standard constructions for practical channels. It is important to note that the beamspace channels in our simulations include leakage effects and are approximately sparse.
\section{Compressive codebook design with perfect AoD prior}\label{sec:CS}
In this section, we assume the availability of a known AoD prior and design an optimization technique that optimizes CS matrix based on the AoD prior using a 2D-CCS framework \cite{myers2019falp}. This optimized CS matrix can be used for more efficient channel measurements to improve the beam alignment performance. In Section \ref{sec:onlinesensing}, we will propose an online learning framework that learns such AoD prior and optimizes CS matrix accordingly. 
\subsection{2D-convolutional CS: Motivation and Background}\label{sec:2dccs}
Channel measurements in 2D-CCS are the projections of the channel matrix onto \emph{2D circulantly shifted} versions of a \emph{base matrix} $\rP$ \cite{myers2019falp}. For $M$ channel measurements with 2D-CCS, the BS applies $M$ distinct 2D-circulant shifts of $\rP$ to its phased array. The set of the circulant shifts used to acquire the $M$ measurements is denoted as $\Omega$, given by 
\begin{align}
\Omega = \{(r[1], c[1]), (r[2], c[2]), \cdots, (r[M], c[M])\}.
\end{align}
The coordinates $\{(r[m], c[m])\}_{m=1}^{M}$ are chosen at random without replacement from $[N]\times [N]$. In the $m$-th training slot, the phase shift matrix $\rP[m]$ in 2D-CCS is obtained by circulantly shifting the base matrix $\rP$ by $r[m]$ rows and $c[m]$ columns. The receiver acquires the projection of $\rH$ on $\rP[m]$ according to \eqref{eq:sysmodel}. The motivation to use 2D-CCS is two fold. First, the CS matrix in 2D-CCS can be parametrized just by the base matrix $\rP$ and the set $\Omega$. Such a parametrization results in fewer optimization variables compared to that in generic CS matrix optimization. Second, 2D-CCS can exploit the AoD prior very well to design a proper base matrix suited to site-specific layouts of the streets. As the magnitude of the 2D-DFT of a matrix is invariant to its 2D-circulant shifts \cite{kak2002principles}, the 2D-DFT magnitude of all the phase shift matrices in 2D-CCS is exactly the same. The 2D-DFT magnitude of a phase shift matrix represents its beam pattern sampled at discrete angular locations. A reasonable channel measurement strategy under a AoD prior is to use beam patterns that focus more power along the directions that are more likely to have the strongest components in the channels. 

\par Now, we explain the mathematical idea underlying 2D-CCS. We define $\rP_{\mathrm{FC}}$ as the flipped and conjugated version of $\mathbf{P}$, i.e., $\mathbf{P}_{\mathrm{FC}} (k,\ell)= \overline{\mathbf{P}}(N-k,N-\ell)\;\; \forall k, \ell$. To demonstrate how 2D-CCS works, we obtain the 2D circular convolution of the channel and $\rP_{\mathrm{FC}}$ as
\begin{align}\label{equ:allmeasurement}
\rG = \rH\circledast \rP_{\mathrm{FC}}.
\end{align}
The $k, \ell$-th entry $\rG(k, \ell)$ is the projection of the channel onto the phase shift matrix obtained by shifting $\rP$ by $k$ rows and $\ell$ columns. We use $\mathcal{P}_{\Omega}(\cdot):\mathbb{C}^{N\times N} \to \mathbb{C}^{|\Omega| \times 1}$ to represent the projection operator that returns the entries of a matrix at the locations in $\Omega$ as a vector. We define the noise vector $\rv\in\mathbb{C}^{|\Omega|\times 1}$. With this definition, the channel measurements in 2D-CCS can be expressed as 
\begin{align}
\label{eq:basic_subsample}
 \ry =\mathcal{P}_{\Omega}(\rH\circledast \rP_{\mathrm{FC}}) + \rv.
 \end{align}
 Therefore, it can be observed that the channel measurements in 2D-CCS are a subsampling of the convolution between $\rH$ and $\rP_{\mathrm{FC}}$. 
\par We use the mask concept in \cite{myers2019falp} to transform the CS problem corresponding to (\ref{eq:basic_subsample}) into a partial 2D-DFT CS problem. The transformation will be used to simplify our CS matrix optimization problem. Based on the DFT property, the inverse 2D-DFT of $\rG$ is given by \cite{kak2002principles}
\begin{align}\label{equ:DFT2Dcirc}
\rU_N^* (\rH\circledast\rP_{\mathrm{FC}})\rU_N^*  = \underbrace{\rU_N^*\rH\rU_N^*}_{\rX} \odot \underbrace{{N}\rU_N^*\rP_{\mathrm{FC}}\rU_N^*}_{\rZ}, 
\end{align}
which is represented as the Hadamard product of two terms, $\rX$ and $\rZ$. 
The left side of the Hadamard product in (\ref{equ:DFT2Dcirc}) is the channel beamspace. The right side of the Hadamard product is defined as the \emph{mask}. As $\Vert \rP \Vert _{\mathrm F}=1$, it can be observed that $\Vert \rZ \Vert _{\mathrm F}=N$. We name the Hadamard product of $\rX$ and $\rZ$ as the masked beamspace \cite{myers2019falp}. It can be observed from \eqref{equ:DFT2Dcirc} that $\rH\circledast\rP_{\mathrm{FC}}=\rU_N(\rX \odot \rZ)\rU_N$. We substitute the 2D-DFT representation of (\ref{equ:DFT2Dcirc}) in \eqref{eq:basic_subsample}, which gives
 \begin{align}
 \ry =\mathcal{P}_{\Omega}\left(\rU_N(\rX\odot \rZ)\rU_N\right) + \rv.\label{equ:XotimeZ}
 \end{align}
As the masked beamspace $\rX \odot \rZ$ is sparse, it can be recovered from its subsampled 2D-DFT in  (\ref{equ:XotimeZ}) with standard partial 2D-DFT techniques. To guarantee successful recovery of the beamspace $\rX$, however, the mask $\rZ$ should be non-zero at all locations. To satisfy such non-zero constraints, a \emph{unimodular} mask was used in \cite{myers2019falp}. Denoting the estimated beamspace as $\hat{\rX}$, it was shown that the use of a unimodular mask minimizes the channel reconstruction error $||\rX - \hat{\rX}||_\mathrm{F}$ \cite{myers2019falp}.
 
Non-uniform masks can be more efficient to acquire compressive channel measurements when the AoD prior is known at the BS. As the AoD prior in vehicular settings follows a certain statistical pattern according to the street layouts, designing a mask $\rZ$ that is tailored to the prior can result in better beam alignment. In this paper, we define the AoD prior as a probability distribution of the best beam index within the 2D-DFT beamspace. A good mask is one with a larger magnitude along the directions that are more likely to be optimal. As the squared Frobenius norm of the mask is constrained to be $N^2$, the mask design problem is analogous to a power allocation problem. In Section \ref{sec:nonuniformsensing}, we investigate the mask design problem for a given AoD prior.
 \subsection{Mask design with AoD prior}\label{sec:nonuniformsensing}
CS matrices in 2D-CCS are parameterized by $\rP$, or the mask $\rZ$, and the subsampling set $\Omega$. In this paper, we focus on the optimization over the magnitude of the mask and leave optimization of the  subsampling set $\Omega$ to future work. To this end, we assume $\Omega = [N] \times [N]$. Under such full sampling assumption, (\ref{equ:XotimeZ}) can be rewritten as 
\begin{align}
\label{eq:fullsamp_meas}
\ry =\mathrm{vec}\left(\rU_N(\rX\odot \rZ)\rU_N \right) + \rv.
\end{align}
We define $\tilde{\ry}$ as an $N^2 \times 1$ vector that is obtained by inverting the 2D-DFT in \eqref{eq:fullsamp_meas}, i.e., $\tilde{\ry}=\left(\rU^{\ast}_N \otimes \rU^{\ast}_N\right) \ry$. We define the transformed noise vector as $\tilde{\rv}=\left(\rU^{\ast}_N \otimes \rU^{\ast}_N\right) \rv$ to write 
\begin{align}
\tilde{\ry} & = \vec(\rX \odot \rZ) + \tilde{\rv}\\
& = \underbrace{\vec(\rX)}_{\rx}\odot \underbrace{\vec(\rZ)}_{\rz} + \tilde{\rv}.
\end{align}
\iffalse 
To formulate the problem, we first assume that CS implements \emph{full} sampling of the channel, i.e., $\Omega = [N] \times [N]$. Based on the full sampling assumption, we write  (\ref{equ:XotimeZ}) as
where (\ref{equ:unitary}) is obtained due to the unitary-invariant property of the DFT matrix $\rU_N$. By applying vectorization to (\ref{equ:byfullsampling}), we have 
\begin{align}
\tilde{\ry} & = \vec(\rX \odot \rZ) + \tilde{\rv}\\
& = \underbrace{\vec(\rX)}_{\rx}\odot \underbrace{\vec(\rZ)}_{\rz} + \tilde{\rv},
\end{align}
\fi
Due to the unitary nature of the 2D-DFT, $\tilde{\rv}$ has the same statistics as $\rv$, i.e., $\tilde{\rv}\sim\mathcal{N}_\mathrm{c}(0, \sigma^2 \rI)$. We assume that the one-sparse beamspace vector $\rx$ is an instance of a random variable $\boldsymbol{\mathsf{x}}$ that is supported on a discrete event set $\mathcal{F}$. Under the one-sparse prior assumption, $\mathcal{F}$ contains the canonical basis vectors $\mathbf{e}_1, \mathbf{e}_2, \cdots, \mathbf{e}_{N^2}\in\mathbb{C}^{N^2 \times 1}$. The probability that $\boldsymbol{\mathsf{x}}=\boldsymbol{\mathsf{e}}_k$, i.e., the $k^{\mathrm{th}}$ beamspace component is $1$, is $p_k$. The AoD prior is defined by the probabilities $\{p_k\}_{k=1}^{N^2}$ that sum to one. The event that $\boldsymbol{\mathsf{x}}=\boldsymbol{\mathsf{e}}_k$ corresponds to the case in which the $k$-th beam direction is optimal in the 2D-DFT beamspace. 
\par Now, we derive the probability of successful beam alignment for a mask $\rz$. For such a computation, we first derive the success probability conditioned on $\boldsymbol{\mathsf{x}}=\mathbf{e}_k$. When the $k^{\mathrm{th}}$ beamspace direction is optimal, the entries of $\tilde{\ry}$ can be expressed as 
\begin{align}\label{equ:receivesignal}
\tilde{\y}_k &=1\cdot {\z}_k+ \tilde{\mathrm{v}}_k={\z}_k+ \tilde{\mathrm{v}}_k, \\
\tilde{\y}_j&= 0\cdot {\z}_j+ \tilde{\mathrm{v}}_j= \tilde{\mathrm{v}}_j, ~\forall j \neq k.
\end{align}
For $\boldsymbol{\mathsf{x}}=\mathbf{e}_k$, beam alignment is successful when $|\tilde{\y}_k|>|\tilde{\mathrm{y}}_j|, ~\forall j \neq k$, i.e.,
\begin{align}
|\z_k + \tilde{\mathrm{v}}_k|>|\tilde{\mathrm{v}}_j|, ~\forall j \neq k.
\end{align}
Such an event occurs with a probability $\prod _{j \neq k} \mathbb{P}\left(|\tilde{\mathrm{v}}_j|<|\z_k + \tilde{\mathrm{v}}_k|\right)$. Considering the $N^2$ candidates for $\boldsymbol{\mathsf{x}}$, the probability of successful beam alignment is
\begin{align}
\mathbb{P}(\text{successful alignment}) &\mathop{=}^{(a)} \sum_{k = 1}^{N^2} p_k \prod _{j \neq k} \mathbb{P}\left(|\tilde{\mathrm{v}}_j|<|\z_k + \tilde{\mathrm{v}}_k|\right)\\
&\mathop{=}^{(b)} \sum_{k = 1}^{N^2} p_k\left(\mathbb{P}\left(|\tilde{\mathrm{v}}| < |\z_k+ \tilde{\mathrm{v}}_k|\right)\right)^{N^2 - 1}\\
& = \sum_{k = 1}^{N^2} p_k\left(\mathbb{P}\left(\bigg|\frac{\tilde{\mathrm{v}}}{\z_k}\bigg|< \bigg|1 + \frac{\tilde{\mathrm{v}}_k}{\z_k}\bigg|\right)\right)^{N^2 - 1}, \label{equ:suc_align}
\end{align}
where $(a)$ is derived based on the assumption that the events in $\mathcal{F}$ are independent, and $(b)$ follows from the IID noise assumption.
\par We express the probability of successful beam alignment as a function of the AoD prior, the mask, and the noise variance $\sigma^2$. Conditioned on $\boldsymbol{\mathsf{x}}=\mathbf{e}_k$, the random variable $\frac{\tilde{\mathrm{v}}}{\z_k}\sim \mathcal{N}\left(0, \frac{\sigma^2}{|\z_k|}\right)$.We use Lemma \ref{lemma:prob} to simplify \eqref{equ:suc_align}. 
\begin{lemma}\label{lemma:prob}
For $\mathsf{x}$ and $\mathsf{y}$ as IID complex Gaussian with $\mathsf{x}, \mathsf{y}\sim \mathcal{N}\left(0, \xi^2\right)$, we have 
\begin{align}\label{equ:prob_lemma}
\mathbb{P}\left(|1 + \mathsf{x}|^2 \geq |\mathsf{y}|^2\right) = 1 - \frac{1}{2}\exp\left(-\frac{1}{2\xi^2}\right). 
\end{align}
\begin{proof}
For convenience of notation, we first rewrite $\mathbb{P}\left(|1 + \mathsf{x}|^2 \geq |\mathsf{y}|^2\right)$ as 
\begin{align}
\mathbb{P}\left(|1 + \mathsf{x}|^2 \geq |\mathsf{y}|^2\right) = \mathbb{P}\left(\bigg |\frac{\sqrt{2}(1 + \mathsf{x})}{\xi}\bigg|^2 \geq \bigg|\frac{\sqrt{2}\mathsf{y}}{\xi}\bigg|^2\right)
\end{align}
Since both $\mathsf{x}$ and $\mathsf{y}$ are complex Gaussian, the distribution of $\mathsf{x}_\mathrm{c}^2 = \big|\frac{\sqrt{2}\mathsf{y}}{\xi}\big|^2$ is a chi-square distribution with degree of 2. The random variable $\mathsf{x}_\mathrm{nc}^2 = \bigg |\frac{\sqrt{2}(1 + \mathsf{x})}{\xi}\bigg|^2$ follows a non-centered chi-square distribution of degree 2 with $\lambda = 2\xi^2$ \cite{lancaster2005chi}. The probability in (\ref{equ:prob_lemma}) can be derived as
\begin{align}
\mathbb{P}\left(|1 + \mathsf{x}|^2 \geq |\mathsf{y}|^2\right) &= \mathbb{P}\left(\mathsf{x}_\mathrm{c}^2\leq\mathsf{x}_\mathrm{nc}^2\right) = 1 - \mathbb{P}\left(\mathsf{x}_\mathrm{c}^2\geq\mathsf{x}_\mathrm{nc}^2\right)\\
&=\mathbb{E}_{\mathsf{x}_\mathrm{nc}^2}\left[1 - \exp\left(- \frac{1}{2}\mathsf{x}_\mathrm{nc}^2\right)\right] \mathop{=}^{(a)} 1 - \frac{1}{2}\exp\left(-\frac{1}{2\xi^2}\right),
\end{align}
where $(a)$ is derived based on the moment-generating function of non-centered chi-square distribution with degree $2$ and $\lambda = 2\xi^2$ \cite{lancaster2005chi}. 
\end{proof}
\end{lemma} We use Lemma \ref{lemma:prob}, to rewrite (\ref{equ:suc_align}) as 
\begin{align}\label{equ:prob_align2}
\mathbb{P}(\text{successful alignment}) = \sum_{k = 1}^{N^2} p_k \left(1 - \frac{1}{2}\exp\left(-\frac{|\z_k|^2}{2\sigma^2}\right)\right)^{N^2 -1}.
\end{align}
The goal of this section is to determine the mask $\rz$ that maximizes the probability in \eqref{equ:prob_align2}. 
\par We now derive a convex optimization problem that maximizes an approximation of the beam alignment probability. As $\Vert \mathbf{P} \Vert_\mathrm{F}=1$, the mask must satisfy the norm constraint $\Vert\rz \Vert_2=N$. The mask optimization problem is then
\begin{align}\label{equ:opt_problem}
%\max \quad &\mathbb{P}(\text{successful alignment}) \text{in} (\ref{equ:prob_align2})\\
\max \quad & \sum_{k = 1}^{N^2} p_k \left(1 - \frac{1}{2}\exp\left(-\frac{|\z_k|^2}{2\sigma^2}\right)\right)^{N^2 -1}\\
\textrm{s.t.} \quad &\sum_{k = 1}^{N^2} |\z_k|^2 = N^2.\nonumber
\end{align}
Unfortunately, the optimization problem in (\ref{equ:opt_problem}) is non-convex. Based on Jensen's inequality, the log-equivalent form of the objective in (\ref{equ:opt_problem}) can be lower-bounded by 
\begin{align}
\log\left(\sum_{k = 1}^{N^2} p_k \left(1 - \frac{1}{2}\exp\left(-\frac{|\z_k|^2}{2\sigma^2}\right)\right)^{N^2 -1}\right) &\geq \sum_{k = 1}^{N^2}p_k\log\left(\left(1 - \frac{1}{2}\exp\left(-\frac{|\z_k|^2}{2\sigma^2}\right)\right)^{N^2 -1}\right)\\
&=(N^2 - 1)\sum_{i = 1}^{N^2 - 1}p_k\log\left(1 - \frac{1}{2}\exp\left(-\frac{|\z_k|^2}{2\sigma^2}\right)\right).\label{equ:lowbound}
\end{align}
Fortunately, the lower bound in (\ref{equ:lowbound}) is convex in $(|\z_1|^2, |\z_2|^2, \cdots, |\z_{N^2}|^2)^T$, i.e., $|\rz|^2$. The constraint in (\ref{equ:opt_problem}) is also convex in the squared magnitude vector $|\rz|^2$. The optimal mask $|\rZ|_\mathrm{opt}$  that maximizes the lower bound on the probability of successful beam alignment can be obtained by the following optimization problem 
\begin{align}\label{equ:opt_problem_approx}
 \max \quad & \sum_{k = 1}^{N^2 - 1}p_k\log\left(1 - \frac{1}{2}\exp\left(-\frac{|\z_k|^2}{2\sigma^2}\right)\right)
\\
\textrm{s.t.} \quad &\sum_{k = 1}^{N^2} |\z_k|^2 = 1.\nonumber
\end{align}
The problem in \eqref{equ:opt_problem_approx} can be solved using standard convex optimization by considering $\{|\z_k|^2\}^{N^2}_{k=1}$ as the variables. We can observe that (\ref{equ:opt_problem_approx}) shares similar structure to the classical \emph{waterfilling} problem, by approximating the exponential function in (\ref{equ:opt_problem_approx}) to its linear approximation. In this paper, we do not make such an approximation. Instead, we use the formulation in (\ref{equ:opt_problem_approx}) to find the mask magnitude that is optimized for an AoD prior. We use the \texttt{CVXPY} package in Python  \cite{diamond2016cvxpy} to solve the convex optimization problem in (\ref{equ:opt_problem_approx}).
\subsection{Phase recovery from the optimized mask magnitude}\label{sec:phaseretrieval}
From (\ref{equ:DFT2Dcirc}), it can be observed that the phase shift matrix $\rP$ is the inverse 2D-DFT of the mask $\rZ$. The optimization problem in (\ref{equ:opt_problem_approx}), however, only provides the magnitude of the mask. It is important to note that the phase of the mask cannot be arbitrary. This is because the base phase shift matrix $\rP$, i.e., the 2D-DFT of $\rZ$, must satisfy the unit-norm constraint of $|\rP|=\mathbf{1}/N$. The problem of finding a phase of a matrix such that its inverse 2D-DFT has a known magnitude has been investigated in optics \cite{gerchberg1972practical}. The Gerchberg-Saxton algorithm in \cite{gerchberg1972practical} iteratively retrieves the phase of a pair of signals related by Fourier transform, with the availability of the amplitudes of the two signals. In our problem, the base matrix $\rP$ and the mask $\rZ$ form a Fourier transform pair, with known amplitudes, i.e., $\mathbf{1}/N$ and the optimized mask $|\rZ|_\mathrm{opt}$. The BS computes the phase of the mask $\rZ_\mathrm{opt}$ and the corresponding base matrix $\rP_\mathrm{opt}$ by the \emph{Gerchberg-Saxton algorithm} \cite{gerchberg1972practical}. The BS applies $M\ll N^2$ random circulant shifts of the resultant base matrix $\rP_\mathrm{opt}$ for the receiver to acquire 2D-CCS-based channel measurements.
\iffalse 
The Gerchberg-Saxton algorithm can be explained in Algorithm \ref{alg:GerchbergSaxton}.

\begin{algorithm}
 \caption{Gerchberg-Saxton algorithm for phase retrieval.}\label{alg:GerchbergSaxton}
 \begin{algorithmic}[1]
 \renewcommand{\algorithmicrequire}{\textbf{Input:}} 
 \renewcommand{\algorithmicensure}{\textbf{Output:}}
 \REQUIRE Mask amplitude $|\rZ|$, number of iterations $K$.
 \ENSURE Retrieved phase $\mathbf{\Theta}$ of $\rP$, where $\rP = e^{\mathsf{j}\mathbf{\Theta}}$. \\
 \textit{Initialization} : Initialize random phase $\mathbf{\Psi}_0$ of $\rZ$ such that $\rZ = |\rZ|e^{\mathsf{j}\mathbf{\Psi}_0}$. The initial random phases in $\mathbf{\Psi}_0$ is uniformly distributed in $(-\pi, \pi]$. 
  \FOR {$i = 0$ to $K - 1$}
  \STATE $\rF = \texttt{Inverse 2D-DFT}(|\rZ|e^{\mathsf{j}\mathbf{\Psi}_i})$.
  \STATE $\rP = e^{\mathsf{j}\mathrm{phase}(\rF)}$.
  \STATE $\mathbf{\Psi}_{i + 1} = \mathrm{phase}(\rP)$.
  \ENDFOR
 \RETURN $\rP$ 
 \end{algorithmic} 
 \end{algorithm}
 
 \fi
  \section{Online sensing matrix learning}\label{sec:onlinesensing}
In Section \ref{sec:CS}, we explained how to design the sensing matrix in 2D-CCS with perfect AoD prior. Such prior can be obtained from a large number of channel realizations measured over a long time, which is similar to the procedure of \emph{offline learning}. The establishment of such offline dataset, however, may not be easy. First, to identify the optimal beam index, a significant amount of overhead will be introduced due to the deployment of large antenna arrays and complicated channel statistics. Second, it is not clear how many data samples would be sufficient to estimate  AoD distribution that is \emph{accurate} enough for the design of the sensing matrix. Despite the changes in the traffic flow such as the traffic density and speed, we can tune the frequencies of CS matrix adaptations in different scenarios. In this paper, we show how online learning can be used to establish the angular domain database by learning the AoD distribution \emph{on the fly}. 

\begin{figure}
\centering
\includegraphics[trim = 0cm 2cm 0.5cm 0cm, width=0.6 \textwidth]{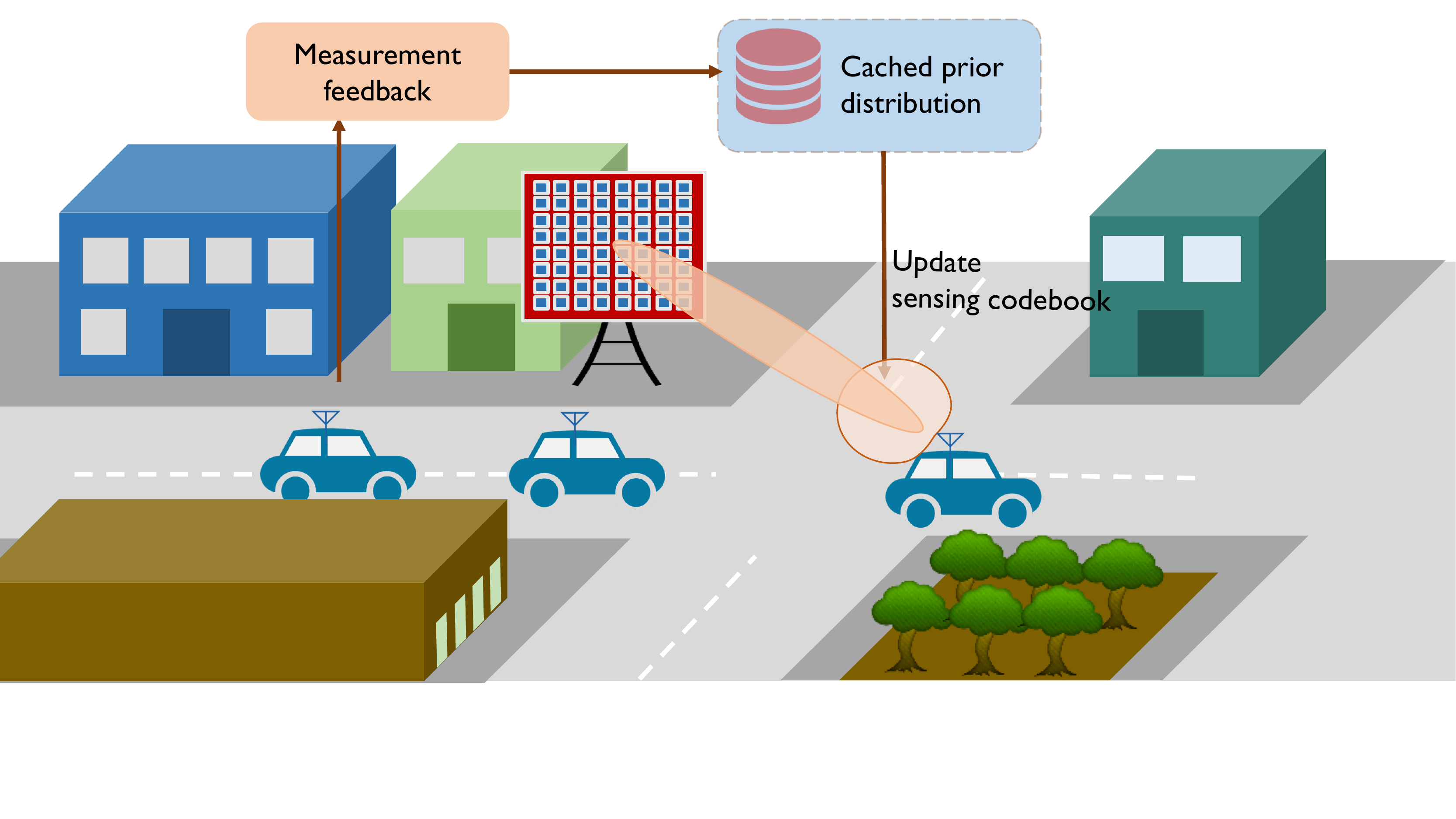}
\caption{An illustration of the online sensing matrix learning. The BS keeps updating and caching an AoD distribution corresponding to the vehicles in its coverage area. Given the available AoD prior, the BS designs its CS matrix for 2D-CCS based on the procedure in Section \ref{sec:CS} and compressively measures the channel. Based on the channel measurement feedback, the BS estimates the direction with the strongest component in the 2D-DFT beamspace and updates the AoD prior accordingly. The BS keeps measuring the channels with a CS matrix optimized based on the available AoD prior and updating such AoD prior iteratively until the estimated AoD distribution converges.  }
\label{fig:illustration}
\end{figure}

Online learning starts with the absence of any useful prior, i.e., the AoD distribution is uniform. The BS constructs the sensing matrix corresponding to the uniform AoD prior according to the procedure in Section \ref{sec:CS}. The receiver feedbacks the CS-based channel measurements to the BS. The BS then recovers the channel beamspace based on the channel feedback and updates the AoD distribution with the estimated optimal beam direction. Fig. \ref{fig:illustration} provides an illustration of the proposed online codebook learning solution.
We use a classical online learning solution in MAB, the UCB algorithm, to update the AoD prior.
\subsection{Multi-armed bandits}\label{sec:UCBbasics}
A MAB is an online learning approach that learns the \emph{policy} of actions and makes decisions from the \emph{rewards} observed so far. The MAB problem can be explained as follows. There is a player with $K$ independent machines, each with a reward distribution that is \emph{initially unknown} to the player. The reward distributions of the $K$ machines are assumed to remain constant at different time steps of the learning. The player plays an action $a_t$ (tries one machine) from the given set $\mathcal{A} = [K]$ and observes the associated reward $R(t)$. The action of the player at time $t$ is only dependent on the past history of observed rewards $H(t - 1) = R(1), R(2), \cdots, R(t - 1)$. A policy $\pi$ is defined as the \emph{mapping} from the history $H(t - 1)$ to the action $a(t)$. \iffalse 
Denoting the mean of $k$-th machine's reward as $\mu(k)$, the regret is defined as the expectation of the difference between the cumulative reward obtained by playing the best arm and the total reward obtained by using the algorithm (or policy) \cite{gittins2011multi}. The regret of the $k$-th machine at time $T$ is defined as 
\begin{align}
\mathbb{R}_T(\pi, k) = T\max_{k \in [K]} \mu(k) - \sum_{t = 1}^T R(t).
\end{align}
A MAB aims at minimizing the expected \emph{regret} over a total of $T$ rounds. 
\fi When the MAB algorithm becomes fully \emph{confident} about its learned policy, it selects the machine that has the largest {expected reward} afterwards.

Online learning algorithms require a careful design of the \emph{exploration} and \emph{exploitation}. There exists a tradeoff between: 1) exploring the environment to find profitable actions, and 2) taking the empirically best action as often as possible. The tradeoff arises because only a limited number of machines are played at each time slot. The algorithm may repeat playing the optimal machine it observes so far from the rewards. Hence, the algorithm has to explore the environment to improve
its knowledge about the reward generating process. \iffalse 
The exploration, however, does not necessarily maximize
the current reward.
\fi In UCB algorithms, such exploration-exploitation status is quantified by \emph{upper confidence index} associated with each machine \cite{gittins2011multi}. \iffalse 
The calculation of the upper confidence index, however, is not intuitive. Not only does the index rely on the entire rewards sequence received so far for each machine, but it also needs to properly \emph{quantify} the extent to which each machine has been explored. \fi In  \cite{auer2002finite}, a family of policies was introduced where the index can be expressed as simple function of the sequence of the actions and rewards of each machine. 

In UCB1 algorithm \cite{auer2002finite}, for example, the upper confidence index is the sum of two terms: 1) the empirical mean of the rewards obtained from the sequence of observations, and 2) the \emph{width} of the confidence bounds that reflects the uncertainty of the player's knowledge. In the basic setting, we assume a $K$-armed bandit problem \cite{gittins2011multi}. We denote the time that arm $k$ has been selected till time $t - 1$ as $T_k(t - 1)$. The time slots in which arm $k$ is played are $\tau^k_1, \tau^k_2, \cdots, \tau^k_{T_k(t - 1)}$. The empirical mean of the reward of $k$-th arm at time $t $ is $\hat{\mu}_k(t - 1) = \frac{1}{T_k(t - 1)}\sum_{j = 1}^{T_k(t - 1)} R(\tau^k_j)$. For UCB1 algorithm, the upper confidence index $\texttt{UCB}_k$ of the $k$-th machine at time $t$ is defined as 
\begin{align}\label{equ:UCB}
\texttt{UCB}_k(t, \delta) = \begin{cases} \infty, &\quad \text{if}~ T_k(t - 1) = 0\\
\underbrace{\hat{\mu}_k(t - 1)}_{\text{Empirical mean}}+ \underbrace{\sqrt{\frac{2\log(1/\delta)}{T_k(t - 1)}}}_{\text{Exploration term}}, &\quad \text{if}~ T_k(t - 1)>0
\end{cases}.
\end{align}
In each time step, the player plays the arm with the largest UCB value. \iffalse If there are multiple arms with the same largest UCB, the player will simply select one of these arms randomly. \fi
\iffalse 
If an arm has never been played till time $t$, i.e., $T_k(t - 1) = 0$, the policy forces the player to choose the unplayed arm whose UCB value is infinity. For the arm that has been played before with $T_k(t - 1)>0$, the UCB value is the sum of two terms as shown in (\ref{equ:UCB}). \fi The first term quantifies the knowledge the player accumulates so far as the empirical mean. The second term evaluates exploration as the square root of the inverse of the time the arm has been played.  The algorithm suppresses the exploration for arms that have been tested many times, while \emph{encourages} those arms that have been tried for only few times. The parameter $\delta$ can be used to control the extent of the exploration \cite{book:banditalg}. When time $t\to\infty$, $T_k(t)\to\infty, ~\forall k$ and the exploration term in (\ref{equ:UCB}) becomes zero. The UCB value becomes the empirical mean and converges to the expected rewards of the machines. The player simply chooses the arm with the largest expected reward at the end of the learning.
\subsection{Online sensing matrix learning with UCB}\label{sec:sensingUCB}
The UCB algorithm fits the online beam alignment problem using a beam scanning-based approach, but its application to the online sensing matrix learning problem is not straightforward. For exhaustive beam search, there exists a finite-sized beam codebook, where the beam codewords can be treated as machines \cite{va2017inverse}. The BS can specify which beam direction it wants to measure by simply applying the beam codeword corresponding to the desired beam direction. In our CS-based framework, the BS obtains an approximation of the whole beamspace using compressed channel measurements. \iffalse Hence, the notion of \emph{machine} is not straightforward.  The beam direction that is updated in CS might not be in accordance with the beams that are \emph{expected} to be measured. \fi
Consider an example where the BS wants to test the $k$-th beam direction. Instead of obtaining feedback of the beamspace $\x_k$ directly through beam search, the BS recovers an estimate of the whole beamspace $\hat{\rx}$ with CS. The \emph{mismatch} between the estimated beamspace and the true beamspace, i.e., $\hat{\rx} - {\rx}$,  could pose great challenges to the exploration-exploitation design in online learning. The ultimate goal of the paper is to learn the sensing matrix that is best suited for a certain vehicular context.
\iffalse  In the proposed model, the reward of each beam direction is a binary value that indicates if the beam direction is optimal. \fi
\iffalse 
The procedure is as follows: (1) The BS initializes the AoD probability prior as uniform. (2) Based on such prior, the BS calculates and applies the 2D-circulant shifts of the base phase shift matrix at the transmitter. (3) The receiver measures the projections of the channel on the phase shift matrix, feedbacks the channel measurements. 4) The BS recovers the estimated optimal beam direction based on the channel feedback. (5) The BS updates the AoD distribution based on receiver feedback. (6) Repeat (2) - (5).
\fi Since both the prior and the sensing matrix are stored and applied at the BS, the online codebook learning model introduces negligible signaling overhead and fully leverages the computation resources at the BS cloud. 

\subsection{Exploration-exploitation in online sensing matrix learning}\label{sec:eetradeoff}
The online sensing matrix learning problem requires an elaborate design of the exploration-exploitation strategy. Without the availability of any useful prior, the online learning framework starts with a uniform mask. If the exploration term in (\ref{equ:UCB}) is not carefully designed, the BS may apply a uniform mask that does not fully leverage the empirical AoD distribution it learns so far. The use of a uniform mask can lead to poor channel estimation performance and slow convergence of the AoD prior to its real distribution. If the BS uses an AoD distribution that is estimated without sufficient exploration, it can end up using masks that may not have good magnitudes along unexplored directions that are likely to be optimal. The estimated AoD distribution could be more and more \emph{biased} towards the beam directions that are estimated as optimal in the beginning of the learning. We propose three approaches to achieve the exploration-exploitation tradeoff. The first approach targets a smooth update of the AoD prior estimation and a \emph{mild} exploration over different beams by adding an exploration term to the estimated AoD prior. The second approach, instead of imposing an exploration term to AoD estimation, requires the BS to directly apply such exploration term over the mask. Finally, we demonstrate that by changing the number of CS measurements over time, the proposed approach can achieve a more accurate AoD distribution estimation, especially in the beginning of online sensing matrix learning.
\subsubsection{UCB with exploration}\label{sec:ucbexplore}
A good design of an exploration-exploitation tradeoff should guarantee a \emph{smooth} evolvement of UCB values through the learning process. The learning starts with a uniform prior, but gradually learns more about the channel and converges to a \emph{less uniform} AoD distribution that can be used for calculating a better mask. The 2D-CCS solution, however, cannot handle a \emph{peaky} yet wrong prior. The exploitation term in UCB of (\ref{equ:UCB}), which is the empirical mean of the AoD distribution, is susceptible to AoD prior estimation inaccuracy due to an insufficient number of channel realizations in the beginning. The time step $t$ in the online learning is equivalent to the \emph{index} of the sequential BS-vehicle channel realizations.
Since the BS calculates the optimal beam direction based on CS measurements, the reward at time $t$ is an one-hot encoded vector $\rr_t$. With the estimated optimal beam direction at time $t$ denoted as $\ell(t)$, we have $\rr_t[\ell(t)] = 1$ and $\rr_t[k] = 0, \forall k \neq \ell(t), k\in[N^2]$. Under successful CS-based recovery, the BS compressively measuring the channel using 2D-CCS is treated as equivalent to the case where the BS measures all beam directions \emph{exhaustively} and returns the optimal beam direction. Therefore, the BS forms a \emph{superarm} that plays all machines each time, i.e., $T_k(t + 1) = T_k(t) + 1,~\forall i$. With the existence of an identical exploration term for different beam directions in (\ref{equ:UCB}) that decreases over time, the mask is less \emph{sensitive} to the change of AoD distribution estimation.  Hence, the BS can avoid a peaky mask that arises from the inaccurate AoD estimation in the beginning. The slow decrease of the UCB \emph{uniformity} guarantees the smoothness of the AoD prior at certain locations during initial exploration. The proposed algorithm is explained in Algorithm \ref{alg:UCB}. 
\begin{algorithm}
 \caption{UCB for online sensing matrix learning. }\label{alg:UCB}
 \begin{algorithmic}[1]
 \renewcommand{\algorithmicrequire}{\textbf{Input:}} 
 \renewcommand{\algorithmicensure}{\textbf{Output:}}
 \REQUIRE Number of measurements per time slot $M$, a constant $\delta$, number of plays $T$.
     \ENSURE Estimated AoD prior $\hat{\rp} = [\hat{p}_1, \hat{p}_2, \cdots, \hat{p}_{N^2}]$. \\
  \textit{Initialization}: Initialize a uniform AoD prior $\rp = [p_1, p_2, \cdots, p_{N^2}]$, where $p_i =  \frac{1}{N^2}, ~i = 1, 2, \cdots, N^2$. A sufficiently large value is set as $N_\infty = 10000$. \\
  Initialize $T_i(1) = 0, \hat{\mu}_i(t) = 0, ~\forall i$.\\

   \FOR {$t = 0, 1, 2, \cdots$}
  \STATE Compute UCB values $\texttt{UCB}_i(t) \leftarrow \begin{cases} N_\infty, &\quad \text{if}~ T_i(t - 1) = 0\\
\hat{\mu}_i(t - 1) + \sqrt{\frac{2\log(1/\delta)}{T_i(t - 1)}}, &\quad \text{if}~ T_i(t - 1)>0.
\end{cases}$\\
\STATE Update prior $p_i= \frac{\texttt{UCB}_i(t)}{\sum_{j= 1}^{N^2} \texttt{UCB}_j(t)},~\forall i$, and calculate the sensing matrix based on Section \ref{sec:2dccs} - \ref{sec:phaseretrieval}.
\STATE Perform 2D-CCS and return the estimated optimal beam direction index $s$. 
\STATE Update reward and its empirical mean $\rr_t[i] = 0, \hat{\mu}_i(t + 1) = \hat{\mu}_i(t)~\forall i \neq s$,~$\rr_t[s] = 1$, $\hat{\mu}_s(t + 1) = \frac{\hat{\mu}_s(t) T_s(t) + 1}{T_s(t) + 1}$. 
\STATE Update time of being played $T_i(t + 1) = T_i(t) + 1, ~\forall i$.\\
  \ENDFOR
  \STATE$\hat{\rp} = \rp$

 \RETURN AoD distribution estimate $\hat{\rp}$.
 \end{algorithmic} 
 \end{algorithm}
 
 \subsubsection{Mask regularization}\label{sec:mask_reg}

 The second approach applies exploration directly on the mask $\rZ$. In the first approach, we add an \emph{exploration} term to the estimated AoD distribution in (\ref{equ:UCB}) to minimize the error in the AoD prior estimate that can result in a biased mask. In 2D-CCS, the magnitude of the mask represents the power allocated for sensing along different beam directions that are defined by the 2D-DFT dictionary. We propose to directly apply \emph{regularization} over the mask. The BS calculates a mask directly based on the empirical AoD distribution, added by an extra exploration term in (\ref{equ:UCB}) as in Algorithm \ref{alg:UCB}. The algorithm is provided in Algorithm \ref{alg:probest}. It can be observed that Algorithm \ref{alg:UCB} and \ref{alg:probest} share very similar inputs, outputs, and procedures. The difference is where the exploration (regularization) term is imposed on. The regularization is inexplicitly applied for the AoD estimation of UCB in Section \ref{sec:ucbexplore}, while it is directly incorporated in the mask for probability estimation.  
 
 \iffalse 

 we take into account some details in the estimated optimal beam direction in the exploration term. The first approach properly handles the high bias that is introduced to the AoD distribution estimation. As shown in Algorithm \ref{alg:UCB}, there exists a common term of exploration $\sqrt{\frac{2\log(1/\delta)}{T_i(t)}}$, where $T_i(t)$s are identical. In specific, we assume all arms are played at each time step, even though there is only one optimal beam direction that is fed back to the BS and used to update the AoD distribution. When the BS is not confident about the AoD distribution, however, it still wants to explore more about other beams that are not prominent in the available AoD distribution. To capture such statistics of the XX unevenness of the AoD, we propose to add an extra term to exploration to explore beams that have not been estimated as optimal. XXX Add in an extra term in there which is inversely proportional to the estimated AoD, but multiplied by a small very number. 
\fi
\subsubsection{Reducing the number of CS measurements with online learning}\label{sec:measureupdate}
For the exploration-exploitation approaches proposed in Section \ref{sec:mask_reg} and \ref{sec:ucbexplore}, we assume a fixed number of CS measurements applied for each channel realization. Further varying the number of CS measurements is another way to improve AoD estimation accuracy and achieve a tradeoff between exploitation and exploration, without sacrificing too much in terms of overhead. When a larger number of channel measurements are acquired by the receivers,  CS can reveal more accurate information in the beamspace. In the beginning of the online AoD distribution estimation, the BS must use a larger number of CS-based measurements to guarantee successful beamspace recovery. Successful CS-based beam alignment ensures that the BS obtains a more accurate AoD prior. With a better mask, fewer measurements are sufficient to recover the correct the optimal beam direction. Hence, we propose to apply Algorithm \ref{alg:UCB} and \ref{alg:probest} while adapting the number of CS measurements in different \emph{phases} of online learning. We consider the number of measurements as a function of time step $t$ as $M(t)$.
\begin{algorithm}
 \caption{Probability estimation with mask regularization}\label{alg:probest}
 \begin{algorithmic}[1]
 \renewcommand{\algorithmicrequire}{\textbf{Input:}} 
 \renewcommand{\algorithmicensure}{\textbf{Output:}}
 \REQUIRE Number of measurements per time slot $M$, a constant $\delta$, number of plays $T$.
 \ENSURE Estimated AoD prior $\hat{\rp} = [\hat{p}_1, \hat{p}_2, \cdots, \hat{p}_{N^2}]$. \\
   \textit{Initialization}: Initialize a uniform AoD prior $\rp  = [p_1, p_2, \cdots, p_{N^2}]$, where $p_i =  \frac{1}{N^2}, ~i = 1, 2, \cdots, N^2$. A sufficiently large value $N_\infty = 10000$.\\
  Initialize $T_i(1) = 0, \hat{\mu}_i(t) = 0, ~\forall i$.\\
   \FOR {$t = 0, 1, 2, \cdots$}
  \STATE Update the estimated AoD distribution $p_i \leftarrow \frac{\hat{\mu}_i(t)}{\sum_{j = 1}^{N^2} \hat{\mu}_j(t)}$.\\
\STATE Calculate the mask $\rz$ based on $p_1, p_2, \cdots, p_{N^2}$ based on Section \ref{sec:2dccs} - \ref{sec:phaseretrieval}.
\STATE Regularize the mask amplitude with \\
 $|\rz_i|\leftarrow \begin{cases} |\rz_i| + N_\infty, &\quad \text{if}~ T_i(t - 1) = 0\\
|\rz_i| + \sqrt{\frac{2\log(1/\delta)}{T_i(t - 1)}}, &\quad \text{if}~ T_i(t - 1)>0.
\end{cases}$
\STATE
Normalize the mask amplitude $|\rz|$. Apply Gerchberg-Saxton Algorithm as in Section \ref{sec:phaseretrieval} to obtain the phase of the mask. 
\STATE
Perform 2D-CCS with $\rz$ and return the estimated optimal beam direction index $s$. 
\STATE Update reward and its empirical mean $x_i(t) = 0, \hat{\mu}_i(t + 1) = \hat{\mu}_i(t)~\forall i \neq s$,~$x_{s}(t) = 1$, $\hat{\mu}_s(t + 1) = \frac{\hat{\mu}_s(t) T_s(t) + 1}{T_s + 1}$.
\STATE Update the AoD distribution as the empirical mean $p_i= \hat{\mu}_i(t),~\forall i$. 
\STATE Update time of being played for arm $i$, $T_i(t + 1) = T_i(t) + 1, ~\forall i$.\\
  \ENDFOR
  \STATE$\hat{\rp} = \rp$

 \RETURN AoD distribution estimate $\hat{\rp}$.
 \end{algorithmic} 
 \end{algorithm}

The BS uses a higher number of CS measurements at the start of the learning, and decreases $M(t)$ linearly with time. Specifically, $\Delta_t$ defines the time interval after which the CS measurements $M(t)$ is decreased by $\Delta_M$. It should be noted that the number of CS measurements cannot be too small so as to guarantee the accuracy of beam alignment. The initial number of CS measurements is represented as $M_0$, and the lower bound on the number of CS measurements is $M_{\min}$. Hence, $M(t)$ changes over time and can be given by 
    \begin{align}\label{equ:decrease}
        M(t) = \max\left\{M_0 - \floor[\Big]{\frac{t}{\Delta_t}}\Delta_M, M_{\min}\right\}. 
    \end{align}
    Linear decrease of $M(t)$ is a simple way of adjusting measurements temporally. For this approach, multiple hyper-parameters, such as $\Delta_t, ~\Delta_M, ~M_0$, need to be tested and evaluated.
     \section{Numerical results}\label{sec:numerical}
In this section, we evaluate the proposed online 2D-CCS matrix learning framework with comprehensive numerical results. First, we examine the baseline model with perfect AoD prior, i.e., offline learning. Second, we compare performance of the online sensing matrix learning using different exploration-exploitation techniques proposed in Section \ref{sec:eetradeoff}. Last, we demonstrate the convergence of the estimated AoD prior with the proposed online sensing matrix learning algorithms.
\subsection{Offline learning with perfect AoD prior}
\begin{figure}
\centering
\includegraphics[width = 4.0in]{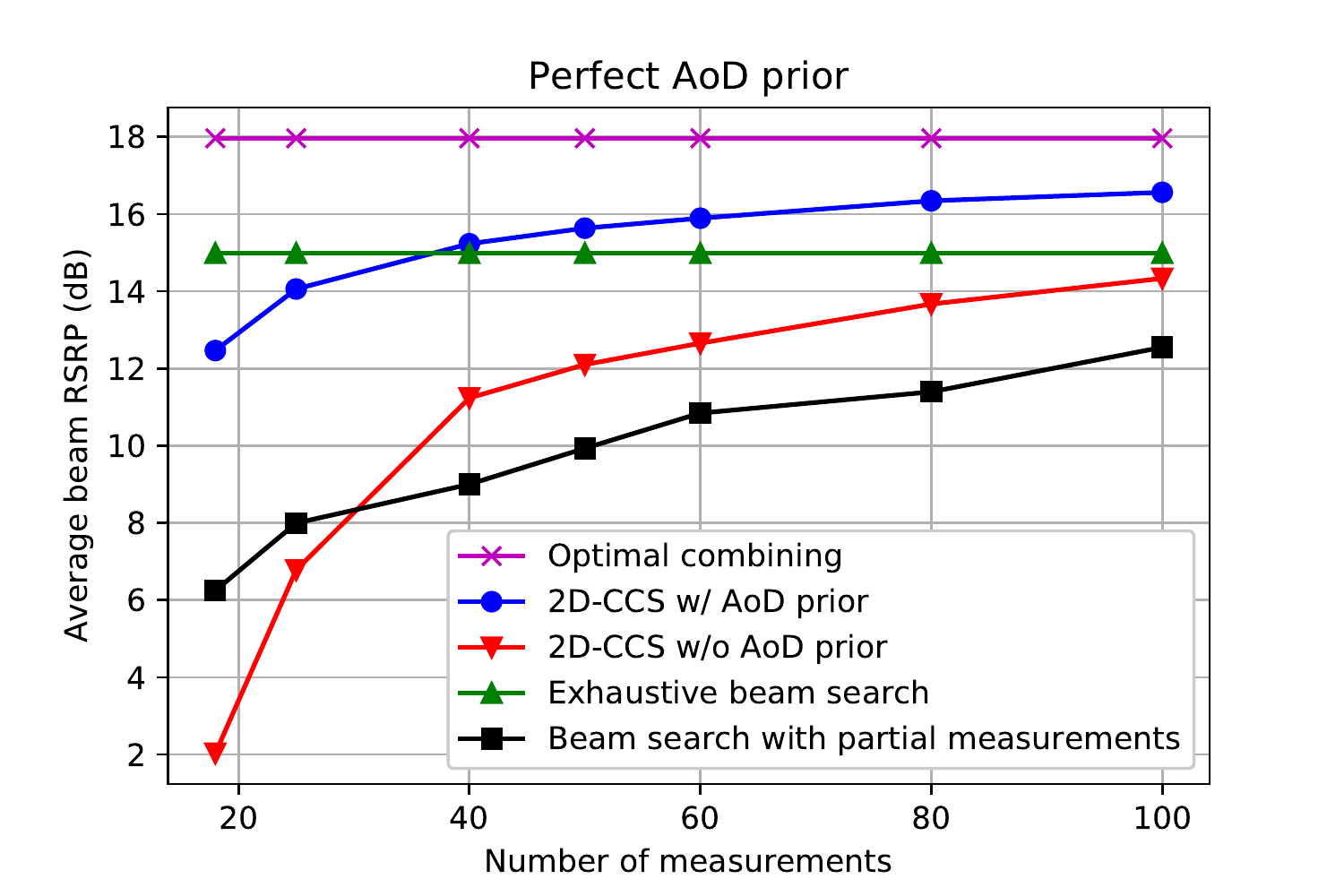}
\caption{A comparison of the average beam RSRP using different beam alignment approaches assuming perfect AoD prior. The proposed 2D-CCS with AoD prior outperforms that using uniform mask by a large margin. The average beam RSRP of 2D-CCS using AoD prior is close to that of optimal beamforming. Furthermore, 2D-CCS with AoD prior exhibits larger beamforming gain compared to exhaustive beam search, when the number of measurements grows large.}\label{fig:perfect}
\end{figure}

In Fig. \ref{fig:perfect}, we evaluate the beam alignment performance of 2D-CCS using perfect AoD prior. The performance metric is the {average beam reference signal received power (RSRP)}. We consider a $32\times32$ UPA at the BS. In the simulation, the transmit power at the BS is adjusted so that the average received SNR for a quasi-omnidirectional
transmission is 10 dB. Such an SNR was achieved by using a Golay code to acquire channel measurements \cite{myers2019falp}. We compare the performance among the proposed 2D-CCS solution, exhaustive beam search with 2D-DFT beam codebook, and the baseline model of optimal beamforming with perfect channel state information (CSI).

The development in this paper assumed a narrowband channel model, but the most interesting applications of mmWave use large bandwidths, and consequently have wideband channels. As a result, we propose a frame structure that allows applying the proposed 2D-CCS-based solution in the wideband case. First, we assume that a complementary Golay pair is used in the time domain for training, which is used in IEEE 802.11ay for example. Exploiting the complementary property, the receiver can separate out the different wideband channel taps. 
There are different ways to apply our algorithm with multiple channel taps $\{\tilde{\rH}_\ell\}_{\ell = 0, 1, \cdots, L - 1}$. Because the channel taps have contributions primarily from a few angular directions, we simply add up the different matrix components. A more sophisticated approach could operate on each tap separately or jointly, but we defer this to future work. 
\iffalse Let tilde y [l] denote the measurement vector .. Our CS ... As the analog ...

In the numerical evaluation, we propose a frame structure that allows
applying the proposed 2D-CCS-based solution in wideband mmWave vehicular communication. We consider $L=64$ taps for the channel in our simulations. For the quasi-omnidirectional reception at the receiver, we denote the $\ell$-th ($0\leq \ell \leq L - 1$) tap of the wideband channel as $\tilde{\rH}[\ell]\in\mathbb{C}^{N\times N}$. In particular, we sum up different taps of channels, i.e., ${\rH} = \sum_{\ell = 0}^{L - 1} \tilde{\rH}[\ell]$ for numerical evaluation. CS-based measurements corresponding to each of the channel taps are obtained using a Golay sequence-based frame structure in \cite{myers2019falp}. The length of the Golay sequence is set to $64$.\fi 

Let $\tilde{\mathbf{y}}[\ell]$ denote the measurement vector corresponding to $\tilde{\mathbf{H}}[\ell]$. Our CS-based online learning framework is applied over the measurement vector ${\ry}=\sum_{\ell = 0}^{L -1}\tilde{\ry}[\ell]$. As the analog beamtraining vectors are frequency flat, CS over $\ry$ is expected to provide an estimate of ${\rH}=\sum_{\ell = 0}^{L - 1} \tilde{\rH}[\ell]$. Let $\rP_\mathrm{BF} $ represent the beamforming matrix applied to the phased array at the BS. The beamforming gain achieved with $\rP_\mathrm{BF}$ is defined as 
\begin{align}\label{equ:bfgain}
\mathrm{BFgain}  = |\langle\rH, \rP_\mathrm{BF}\rangle|^2.
\end{align}
For the perfect CSI case with optimal beamforming,  the beamforming matrix at the phased array is $\rP_{\mathrm{BF},~\text{ideal}}= e^{~\mathsf{j}\mathrm{phase}(\rH)}/N$. We use $\hat{\rH}$ to denote the channel estimate obtained with our online learning-based 2D-CCS algorithm. The orthogonal matching pursuit (OMP) algorithm \cite{tropp2007signal} was used to solve the sparse recovery problem in our method. The beamforming matrix with the proposed approach is  $\rP_{\mathrm{BF,~\text{2D-CCS}}} = e^{~\mathsf{j}\mathrm{phase}(\hat{\rH})}/N$. For exhaustive beam search, the BS searches through the $N^2$ beams from a 2D-DFT beam codebook and identifies the optimal beam direction with the largest beam RSRP. For the beam sweeping among the partial beamspace, with the availability of AoD distribution \emph{ordered} as $\rp = [p_{r_1}, p_{r_2}, \cdots, p_{r_{N^2}}: p_{r_1}> p_{r_2}> \cdots > p_{r_{N^2}}]$, the BS measures the RSRP of the \emph{top} $M$ beams and selects the one with the largest RSRP correspondingly.

Fig. \ref{fig:perfect} shows huge performance improvement with nonuniform mask designed by AoD prior, compared to that with uniform mask in the 2D-CCS framework. With 20 measurements, 2D-CCS with nonuniform mask shows more than 6 dB improvement of average beam RSRP than using beam search with the same number of measurements, and 10 dB  gain compared to 2D-CCS with a uniform mask. With only 35 measurements, 2D-CCS with AoD prior achieves comparable performance with exhaustive search that uses $N^2 = 1024$ beams. The average beam power using 2D-CCS-based beam alignment with AoD prior is superior to exhaustive beam search when $M>40$, and almost achieves the same performance as perfect CSI case. It should be noted that the beam alignment with 2D-CCS outperforms exhaustive beam search because the BS applies conjugate beamforming using the estimated beamspace.
\begin{figure}[h]
\begin{minipage}{0.45\textwidth}
\includegraphics[width = 3.5in]{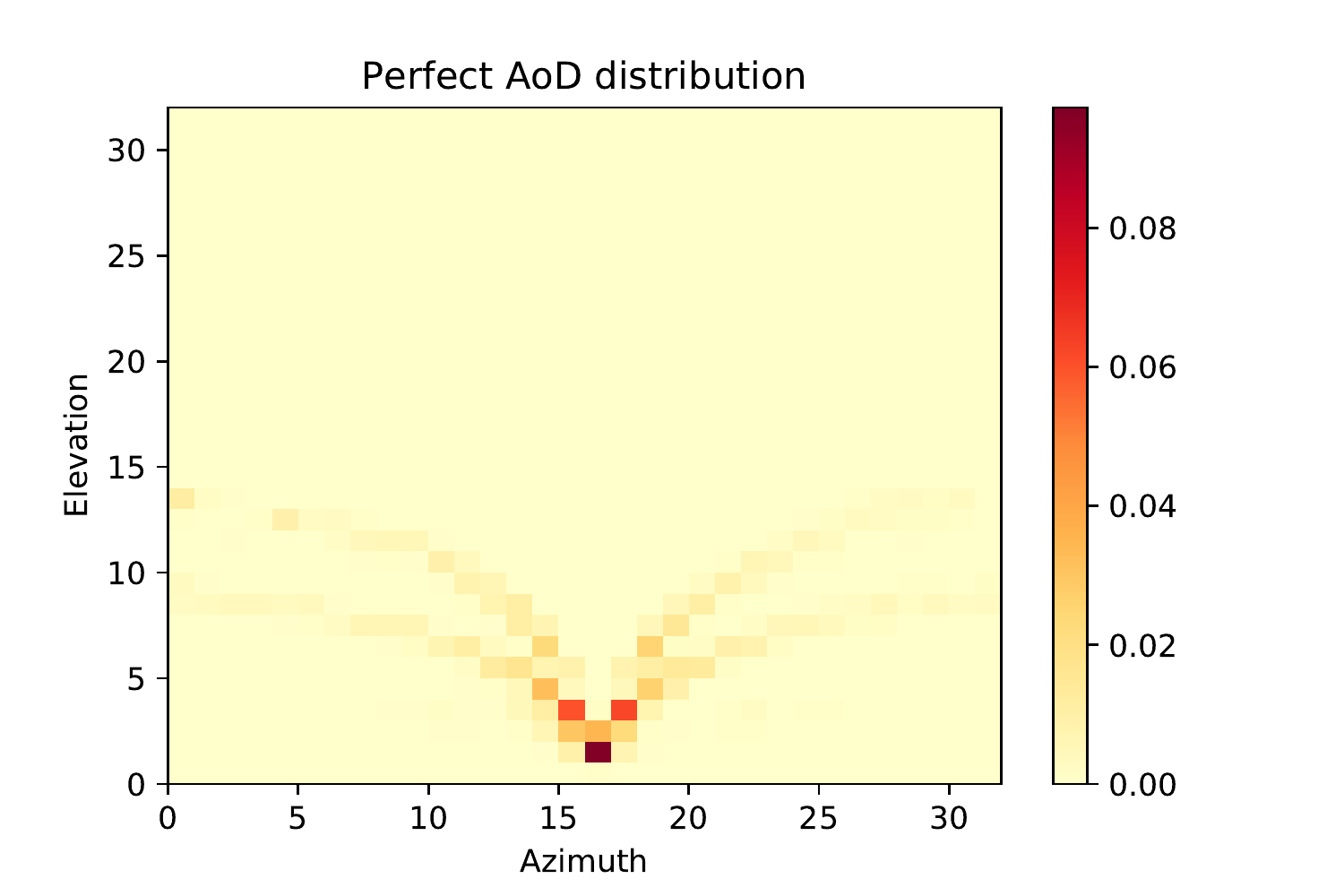}
    \caption{AoD distribution in beamspace across all channel realizations.}\label{fig:realaod}
 \end{minipage}
 \hfill
 \begin{minipage}{0.45\textwidth}
 \includegraphics[width = 3.5in]{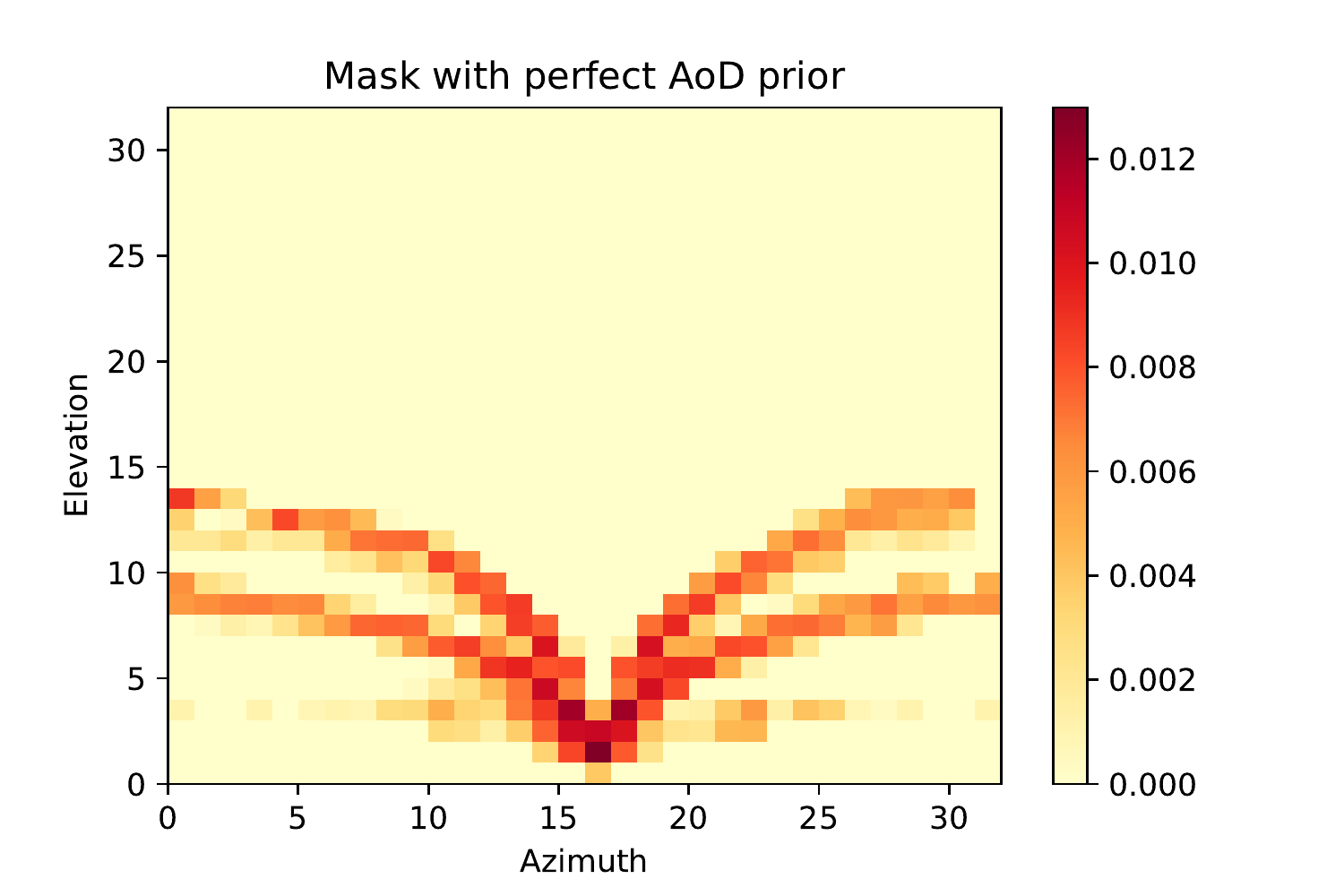}
 \caption{Mask calculated using the optimization procedure in Section \ref{sec:CS} for the AoD prior in Fig. \ref{fig:realaod}}\label{fig:idealmaskaod}
 \end{minipage}
 \end{figure}

 The real AoD distribution and the corresponding mask are plotted in Fig. \ref{fig:realaod} and Fig. \ref{fig:idealmaskaod}. Two \emph{continuous strips} can be observed in Fig. \ref{fig:realaod} and Fig. \ref{fig:idealmaskaod}. These strips represent the two straight lanes in the simulated urban canyon. The observation implies that structured AoD distribution can be extracted from site-specific street layouts and leveraged for CS matrix design. Furthermore, as shown in Fig. \ref{fig:idealmaskaod}, the mask obtained by the optimization in (\ref{equ:opt_problem_approx}) satisfies two desired properties: i) The BS is able to allocate more power to compressively sense the \emph{dominant} beam directions; ii) The mask is also able to cover \emph{less important} beam directions, e.g., the beamspace on the two sides of the strips along the azimuth direction in Fig. \ref{fig:realaod}.   
 \subsection{Online sensing matrix learning}
In this section, we evaluate and compare the effect of \emph{``exploration"} on the online learning performance. We consider the case without any exploration, i.e., directly using the mask calculated by the estimated AoD distribution for sensing, and the three approaches provided in Section \ref{sec:eetradeoff}. 
\subsubsection{No exploration}
\begin{figure}[!h]
    \centering
    \includegraphics[trim = 3.0cm 0cm 0cm 0cm, clip = true,  width = 6.0in]{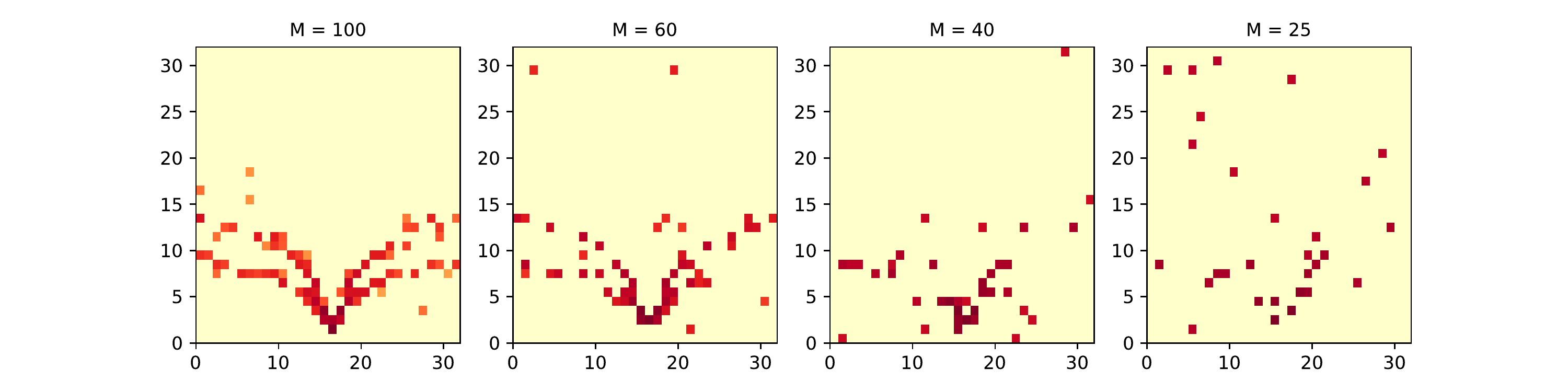}
    \caption{A comparison of the final mask learned without any exploration,  using different numbers of beam measurements $M = 100, 60, 40, 25$. Without exploration, the mask fails to capture the full channel statistics in the angular domain, especially when the number of CS measurements per channel realization is relatively small, e.g., $M = 25$ or $M = 40$.  }
    \label{fig:noexplore}
\end{figure}

Fig. \ref{fig:noexplore} compares the mask learned without any exploration, using a fixed number of measurements. Without exploration, the procedure is identical to Algorithm \ref{alg:UCB} where $\delta = \infty$ and the exploration term is zero. In this case, the sensing matrix is calculated directly based on the empirical AoD distribution without any regularization. As discussed in Section \ref{sec:eetradeoff}, no exploration leads to error propagation and highly biased estimation, which are reflected in the results in Fig. \ref{fig:noexplore}. Compared to the mask calculated from perfect AoD prior, the mask without exploration displays some \emph{inconsistency} in the AoD distribution estimation, and fails to identify several important beam directions. The combined effect of measurement inaccuracy and error propagation due to biased AoD estimation gives rise to a mask that results in poor performance with CS.

\subsubsection{Exploration and regularization}
\begin{figure}
\begin{minipage}{0.43\textwidth}
  \centering
    \includegraphics[width= 3.3in]
    {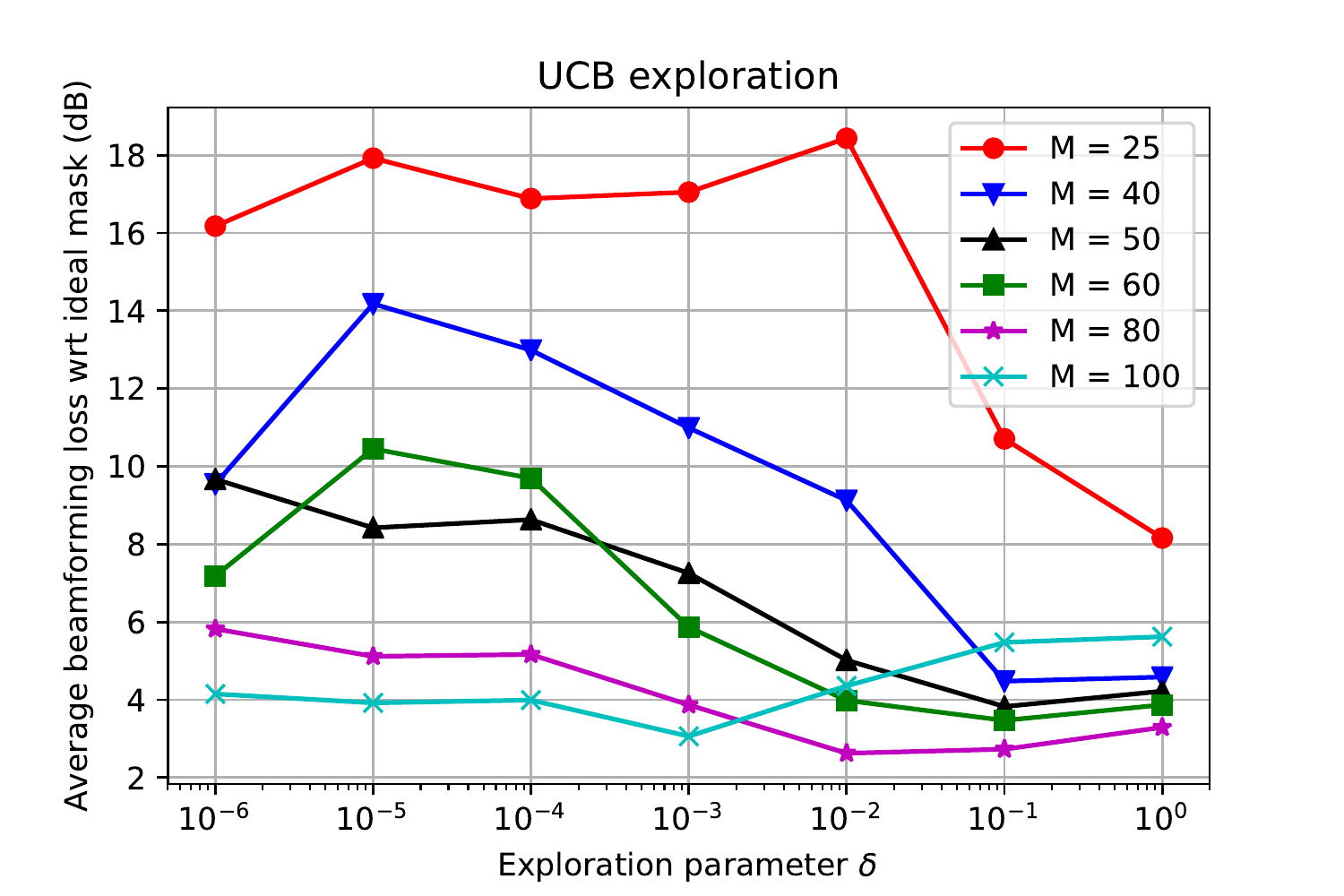}
    \caption{ Comparison of the average beamforming loss using UCB exploration in Section \ref{sec:ucbexplore} with respect to the performance achieved by applying mask in 2D-CCS with perfect AoD prior. }
    \label{fig:fix_M_ucb}
 \end{minipage}
 \hfill
 \begin{minipage}{0.43\textwidth}
 \centering
    \includegraphics[width= 3.3in]
    {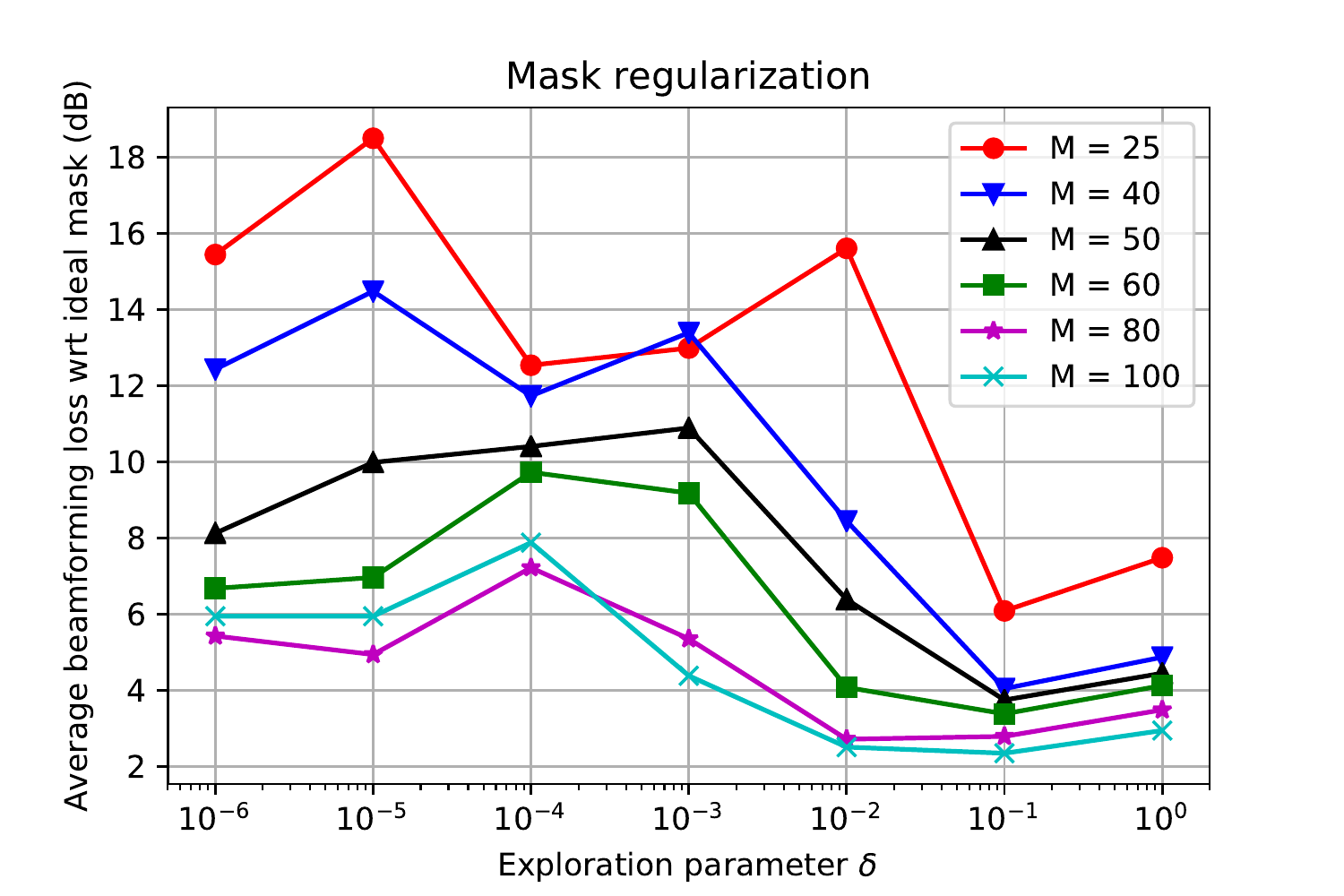}
    \caption{Comparison of the average beamforming loss using mask regularization in Section \ref{sec:mask_reg} with respect to the performance achieved by applying mask in 2D-CCS with perfect AoD prior. }
    \label{fig:fix_M_bf}
 \end{minipage}
 \end{figure}

Fig. \ref{fig:fix_M_ucb} and Fig. \ref{fig:fix_M_bf} illustrate the average beamforming loss with UCB in Section \ref{sec:ucbexplore} and mask regularization in Section \ref{sec:mask_reg}, compared to the \emph{ideal mask} calculated from perfect AoD prior in Fig. \ref{fig:idealmaskaod} using different number of measurements. Specifically, we evaluate the performance of our approach for different values of $\delta$. Fig. \ref{fig:fix_M_bf} shows that the gap between the beamforming gain of the  mask with perfect AoD prior and the one learned online using a fixed number of measurements is marginal, if the regularization parameter $\delta$ is selected carefully. We observe that a low beamforming loss is achieved when $\sqrt{2\log(1/\delta)}$ is around $0.1$. When $\sqrt{2\log(1/\delta)}$  is too large, e.g. $\sqrt{2\log(1/\delta)} = 1$, the mask changes too slowly to capture the information provided by AoD estimation. When $\sqrt{2\log(1/\delta)}$ becomes too small, the learning could be too \emph{aggressive} in the AoD distribution estimation, since it results in a poor mask for 2D-CCS. Similar results of UCB exploration can be observed in Fig. \ref{fig:fix_M_ucb}.
\iffalse For the cases with a large number of measurements, e.g. $M = 60, 80, 100$, however, when $\delta$ is further reduced from $10^{-3}$ to $10^{-5}$, an improvement of the average beamforming gain is shown. The observation is due to the accurate
\fi

Table \ref{tab:mask_reg} provides a more explicit comparison among the beamforming loss of the mask with perfect AoD prior, uniform mask, the mask achieved with UCB exploration, and the mask with mask regularization, with respect to optimal beamforming using perfect CSI. The beamforming loss decreases with a larger number of CS measurements per channel realization. For $M = 40, 50, 60, 80, 100$, the optimal beamforming loss with respect to mask with perfect AoD prior is approximately $1$ dB. When $M = 25$, the regularized mask has superior performance to that of the uniform mask. Also, using only $25$ measurements throughout the whole learning process, the BS is still able to learn the AoD distribution and results in $2.19$ dB beamforming loss compared to the mask calculated using perfect AoD prior. Similar observations can be obtained for different number of CS measurements.

We measure the AoD estimation accuracy based on statistical distance between the estimated AoD distribution and the real AoD distribution of the data samples \cite{mahalanobis1936generalized}.  \iffalse There are multiple measures that can be used to quantify the distance between two probability distribution,  such as the well-known Kullback-Leibler (KL) divergence \cite{van2014renyi}. 
 For two discrete distributions $P = (p_1, p_2, \cdots, p_K)$ and $Q = (q_1, q_2, \cdots, q_K)$, the KL divergence is defined as 
\begin{align}\label{equ:KL}
    \mathcal{D}(P\Vert Q) = -\sum_{i = 1}^Kp_i\log\left(\frac{q_i}{p_i}\right).
\end{align}
 Unfortunately, the KL-divergence in (\ref{equ:KL}) cannot handle zeros in the probability mass functions (PMFs), which is the case for the empirical AoD distribution when only a few beam directions are tried in the beginning. 
 \fi We use a statistical distance measure - Hellinger distance \cite{beran1977minimum}. For two discrete distributions $P = (p_1, p_2, \cdots, p_K)$ and $Q = (q_1, q_2, \cdots, q_K)$, the Hellinger distance $\mathcal{H}(P\Vert Q)$ is defined as 
 \begin{align}\label{equ:hdist}
 \mathcal{H}(P\Vert Q) = \frac{1}{\sqrt{2}}\sqrt{\sum_{i = 1}^{K} (\sqrt{p_i} - \sqrt{q_i})^2}.
 \end{align}
The Hellinger distance is directly related to the Euclidean norm of the difference of the square root of the probability distribution probability mass functions (PMF).

\begin{figure}[!h]
    \centering
    \includegraphics[width = 3.8in]{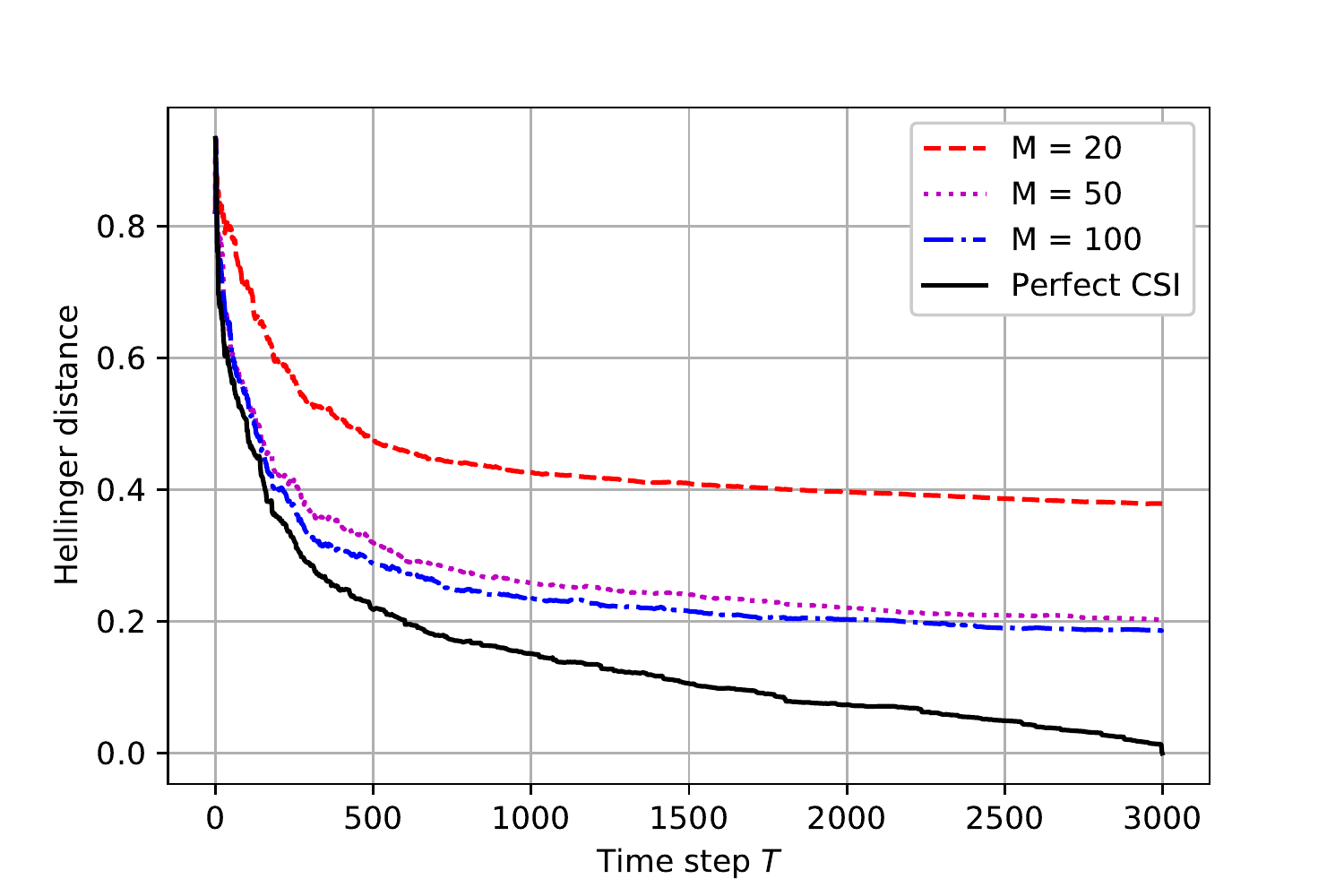}
    \caption{Evaluation of the Hellinger distance defined in (\ref{equ:hdist}) between the real AoD distribution of in total of $3000$ channel realizations, and the empirical AoD distribution learned at different time steps during online sensing matrix learning using mask regularization in Section \ref{sec:mask_reg}. The regularization parameter is set as $\delta = 10$, which is able to achieve a good exploration-exploitation tradeoff as shown in Fig. \ref{fig:fix_M_ucb}
 and Fig. \ref{fig:fix_M_bf}. \iffalse When the number of measurements $M = 50$ and $M = 100$, the achieved AoD estimation accuracies are similar. When applying only $20$ CS measurements per channel realization, the AoD estimation accuracy is relatively low.\fi } \label{fig:hdist}
    \end{figure}

Fig. \ref{fig:hdist} characterizes the temporal evolution of the AoD estimation with mask regularization using a fixed number of measurements. We compare the Hellinger distance between the empirical AoD distribution and the real AoD distribution, with mask regularization, and that with perfect CSI available at the BS. The real AoD prior is calculated across all 3000 channel realizations. The empirical AoD prior is calculated at different time steps, e.g., the AoD prior of the first 200, 300 samples, etc. For the perfect CSI case, we assume the BS can always recover the optimal beam direction. The perfect CSI curve provides a baseline \emph{reference} to evaluate the AoD estimation accuracy \emph{during} online learning. In Fig. \ref{fig:hdist}, the Hellinger distance decreases sharply in the beginning. When time $T<500$, the Hellinger distance with $M = 100, 50$ closely matches that with perfect CSI. After time $T>1000$, the Hellinger distance starts to change slowly. It is also shown that the Hellinger distances of $M = 100$ and $M = 50$ are similar. When $M = 20$, online learning cannot accurately estimate the exact optimal beam direction, which leads to a large Hellinger distance compared to the perfect CSI case.  The performance differences among $M = 100$ and $M = 50$ and using the mask calculated with perfect AoD prior are insignificant, as exhibited in Fig. \ref{fig:perfect} and Table \ref{tab:mask_reg}. The observation implies that the proposed AoD prior-based sensing matrix design is \emph{robust} to small AoD distribution estimation error, which is favorable to real field implementation that generally has more complicated street layouts and channel characteristics.  

\begin{table}
    \centering
     \caption{Comparison of the average BF loss of mask with perfect AoD prior, uniform mask, the mask obtained with UCB exploration, the mask achieved by probability estimation with mask regularization, and the mask with changing number of measurements.}
    \label{tab:mask_reg}
    \begin{tabular}{|c|c|c|c|c|c|c|c|c|}
    \hline
         \diagbox[width=21em]{Avg. beamforming loss \\compared to optimal combining (dB)}{Number of measurements}& $M = 25$ & $M = 40$ & $M = 50$ & $M = 60$ &  $M = 80$ & $M = 100$ \\
         \hline
         \hline
         2D-CCS with perfect AoD prior & 3.90 & 2.73 & 2.33 & 2.07 & 1.62 & 1.40\\ 
         \hline
         2D-CCS with UCB exploration & 8.16 & 4.49 & 3.83 & 3.47 & 2.63 & 3.06 \\\hline
         2D-CCS with regularized mask& 6.09 & 4.05 & 3.75 & 3.39 & 2.72 & 2.35 \\\hline
          2D-CCS with uniform mask & 11.18 & 6.73 & 5.86 & 5.31 & 4.29 & 3.63 \\\hline
         Regularized mask with $M_0 = 200, \Delta_t = 100$  & 5.37 & 3.68 & 3.28 & 3.22 & 2.55 & 2.13\\\hline
          Regularized mask with $M_0 = 200, \Delta_t = 200$ & 5.18 & 3.67 & 3.26 & 3.03 & 2.52 & 2.01 \\\hline
          Regularized mask with $M_0 = 300, \Delta_t = 100$ & 5.07 & 3.56 & 3.38 & 3.00 & 2.47 & 1.96 \\\hline
         Regularized mask with $M_0 = 300, \Delta_t = 200$& 5.21 & 3.62 & 3.41 & 3.13 & 2.50 & 2.05\\\hline
        
    \end{tabular}
   
\end{table}

\subsubsection{Updating the number of CS measurements} The online sensing matrix learning is built on the assumption that the CS measurements can roughly give accurate channel estimation, especially in the beginning of the learning. Starting with a large number of measurements when the BS has no prior is a good way to avoid error propagation in the AoD distribution estimation. After measuring more channel realizations, the BS becomes confident about the AoD distribution estimate. In such a case, decreasing the number of CS measurements can result in a reduced training overhead. With mask regularization, the BS further decreases the number of CS measurements in our online learning-based framework. We present the average beamforming loss with a decreasing number of measurements in the 4 - 7 rows in Table \ref{tab:mask_reg}.  We use $M = M_{\min}$, i.e., the \emph{stabilized} number of measurements the BS uses for each channel realization. In our evaluation, we fix $\Delta_M = 20$. We update $\Delta_t$ in (\ref{equ:decrease}) to modify how fast $M$ decreases. The learning is shown to outperform mask regularization with $M = M_{\min}$. For $M_{\min} = 25$, an extra $1$ dB gain compared to regularized mask can be achieved. No significant differences, however, are observed among the different initial measurement numbers $M_0 = 200, 300$ and the update interval $\Delta_t = 100, 200$. The \emph{worst-case} scenario with the \emph{smallest} initial measurement $M_0 = 200$ and the \emph{fastest} measurement decrease $\Delta_t = 100$ among the four settings in 4 - 7 rows of Table \ref{tab:mask_reg} achieves similar performance compared to the cases with a larger number of measurements $M_0>200$ or $\Delta_t>100$.

\iffalse 
Fig. \ref{fig:entropy} evaluates the change in the entropy of the estimated AoD distribution through the learning using the proposed approaches with different numbers of measurements. The entropy of AoD distribution converges very fast for all approaches, at around $T = 400$. For $M = 20$, Fig. \ref{fig:hdist} shows that the learning cannot exactly capture accurate AoD estimation, which is also reflected in Fig. \ref{fig:entropy}. Using more measurements at the BS and gradually decreasing it, however, makes it possible to establish an accurate prior that can further improve the channel measurement performance afterwards. The initial accurate AoD prior contributes to the smaller beamforming loss -- $1$ dB less loss compared to regularized mask, with only a few measurements, e.g. $M = 20$. 
\begin{figure}
    \centering
    \includegraphics[width = 4.0in]{fig_entropy.pdf}
    \caption{Evaluation of entropy of the empirical AoD distribution at different time steps in the online learning. The entropy of the empirical AoD distribution increases with \emph{accumulated} sampling of the channel realizations over time. The entropy of the empirical AoD distribution converges fast using different number of measurements. }
    \label{fig:entropy}
\end{figure}
\fi
\subsection{Mask convergence}
\begin{figure}
\centering
\includegraphics[trim = 4cm 0cm 4cm 0cm, width = 3.9in]{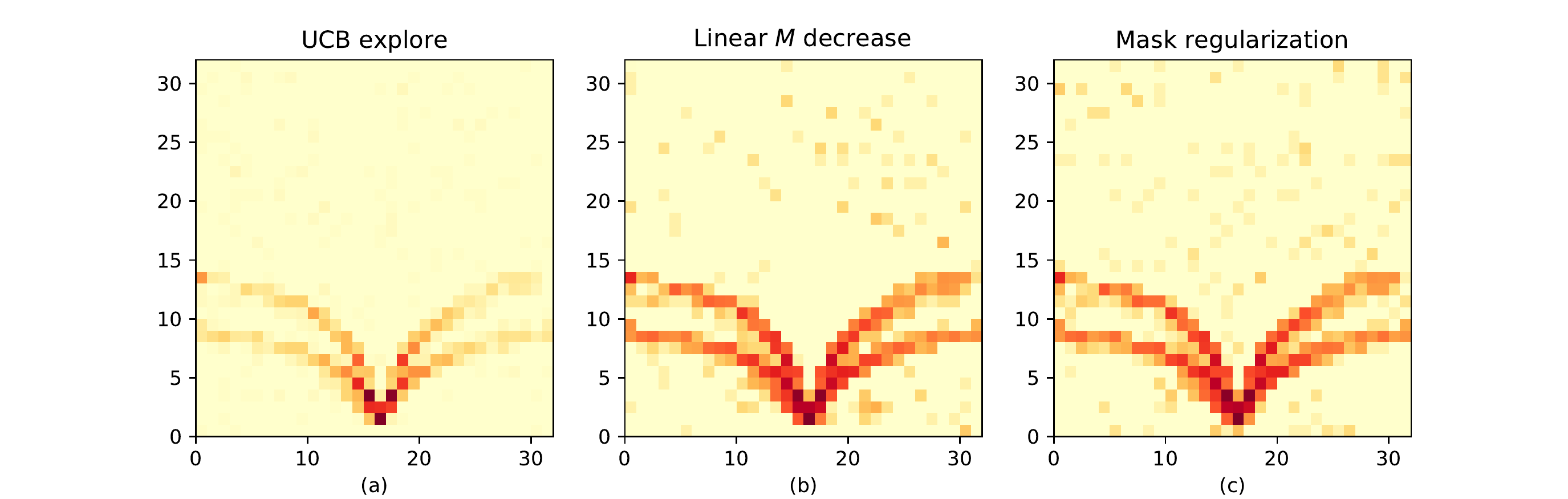}
\caption{Mask convergence for the online learning using UCB exploration, decreasing the number of CS measurements, mask regularization, with $M$ (or $M_{\min}$) = 80.}\label{fig:mask_converge}
\end{figure}

Finally, Fig. \ref{fig:mask_converge} compares the mask convergence of  1) UCB exploration in Section \ref{sec:ucbexplore}, 2) decreasing the number of CS measurements combined with mask regularization in Section \ref{sec:measureupdate}, and 3) mask regularization only in Section \ref{sec:eetradeoff}, after $4000$ time steps. We can observe that all of the three masks can capture the basic structure and the dominant directions of the AoD distribution in the beamspace. For UCB exploration in Fig \ref{fig:mask_converge}.a, the mask fails to capture some \emph{less likely} beam directions. The mask power allocation is more centered around several dominant directions, which makes it difficult to recover a beam direction that has not appeared too often as optimal. \iffalse 
The mask with regularization in Fig. \ref{fig:mask_converge}c, however, exhibits the behavior of \emph{over exploration}. Due to the lack of confidence in AoD estimation in the beginning, the BS directly applies an exploration to the mask which leads to a more uniform mask. With a small number of measurements, it was shown in Fig. \ref{fig:perfect} that uniform mask cannot provide accurate channel estimation. The deficiency of the uniform mask can be compensated by using more channel measurements in the beginning and gradually reducing the number to a predefined value $M_{\min}$ based on (\ref{equ:decrease}). 
\fi The mask convergence for linear measurement number decrease with $M_0 = 300$, $\Delta_M = 100$ is illustrated in Fig \ref{fig:mask_converge}.b. As shown in Fig \ref{fig:mask_converge}.b, the linear measurement decrease procedure achieves better exploration compared to UCB exploration in Fig \ref{fig:mask_converge}.a. In addition, the mask is less \emph{noisy} compared to the converged regularized mask in Fig \ref{fig:mask_converge}.c. The number of CS measurements can be gradually reduced with online learning. The prior learned with our online learning framework enables the BS to perform successful beam alignment with fewer channel measurements than standard CS.
\section{Conclusion}\label{sec:conclusion}
In this paper, we proposed a site-specific  online sensing matrix learning and beam alignment solution in vehicular communication. In a typical vehicular context, the AoD distribution can be structured due to the regularity of the vehicle trajectories and stationary objects such as buildings and roads. Leveraging such AoD distribution statistics in a vehicular scenario, we proposed a novel online compressive sensing framework using a 2D-CCS-based design. We formulated a convex optimization problem that uses the learned AoD prior to design a CS matrix that maximizes the approximate beam alignment probability. We observed superior performance of the AoD prior-aided sensing matrix design compared to either the standard 2D-CCS without AoD prior or exhaustive beam search.  Using only $60$ channel measurements, there was less than $2$ dB gap between the average beam RSRP of the proposed approach and perfect CSI-aided optimal beamforming. 

\iffalse 
We showed good convergence properties of the AoD prior estimation when the BS learns the sensing matrix on the fly. We proposed to leverage a variation of the UCB algorithm, to achieve a tradeoff between exploring more beam directions and exploiting the empirical AoD distribution. We designed three solutions to apply appropriate \emph{regularization} based on different confidence levels of estimation. We showed that online sensing matrix learning achieved comparable performance as offline learning, where perfect AoD prior is available at the BS. Last, we showed that the proposed sensing matrix design is quite robust to small perturbations in estimated AoD distribution. Therefore, online sensing matrix learning is a stronger candidate that adapts to new environment and fits the sequential property of data arrival at BS and user terminals. 
\fi
MmWave vehicular communication has specific channel statistics that need to be exploited for wireless system design. For future work, the design of the \emph{subsampling set} in 2D-CCS that is well-suited to the AoD prior will be investigated. Furthermore, investigating other priors that are based on beam RSRP is an interesting research direction. In addition to the special street layout and specific channel distribution, more information is already embedded in \emph{connected vehicles}. With side information such as location-based situational awareness, the number of measurements can further be reduced if the online learning can reveal the AoD distribution conditioned on the observed side information.

\IEEEpeerreviewmaketitle
\footnotesize
 \bibliographystyle{IEEEtran}
\bibliography{reference}
     \end{document}

%% file: input.tex
\usepackage{amsthm}
\usepackage{etoolbox}
\usepackage{amsmath}
\usepackage{algorithm}
\usepackage{algorithmic}
\usepackage{amsthm} % originally not
\usepackage{url} % originally not
\usepackage{accents} % originally not
\usepackage{amssymb}
\usepackage{amsfonts}
\usepackage{dsfont}
\usepackage[figuresright]{rotating} % originally not
\usepackage{floatpag} % originally not
	\rotfloatpagestyle{empty} % originally not

\usepackage{longtable}% TeXSupport
\usepackage{subfigure}

\usepackage{epstopdf} %originally not
\usepackage{makeidx} %originally not
\usepackage{upgreek} %originally not
\makeindex %originally not

\usepackage{times}
\usepackage{mathdots} % added to allow reverse diagonal dots

\usepackage{bm}

\theoremstyle{plain}% default
\newtheorem{theorem}{Theorem}
\newtheorem{lemma}[theorem]{Lemma}
\AfterEndEnvironment{lemma}{\noindent\ignorespaces}

\def\bb0{{\mathbb{0}}}

\def\bb{{\boldsymbol{b}}}

\def\b0{{\boldsymbol{0}}}

\def\x{{\mathrm{x}}}
\def\y{{\mathrm{y}}}
\def\z{{\mathrm{z}}}
\def\b{{\mathrm{b}}}

\def\rh{{\mathbf{h}}}

\def\rp{{\mathbf{p}}}

\def\rr{{\mathbf{r}}}

\def\rv{{\mathbf{v}}}

\def\rx{{\mathbf{x}}}
\def\ry{{\mathbf{y}}}
\def\rz{{\mathbf{z}}}
\def\r0{{\mathbf{0}}}

\def\rF{{\mathbf{F}}}
\def\rG{{\mathbf{G}}}
\def\rH{{\mathbf{H}}}
\def\rI{{\mathbf{I}}}

\def\rP{{\mathbf{P}}}

\def\rU{{\mathbf{U}}}

\def\rX{{\mathbf{X}}}

\def\rZ{{\mathbf{Z}}}

\def\bsf0{{\bm{\mathsf{0}}}}

\def\N0{{N_{\mathrm{0}}}}

\def\vec{\mathrm{vec}}

\newcommand{\be}{\begin{equation}}
\newcommand{\ee}{\end{equation}}
\newcommand{\bal}{\begin{align}}
\newcommand{\eal}{\end{align}}

%% file: paper_revision_heath_v2.bbl
% Generated by IEEEtran.bst, version: 1.14 (2015/08/26)
\begin{thebibliography}{10}
\providecommand{\url}[1]{#1}
\csname url@samestyle\endcsname
\providecommand{\newblock}{\relax}
\providecommand{\bibinfo}[2]{#2}
\providecommand{\BIBentrySTDinterwordspacing}{\spaceskip=0pt\relax}
\providecommand{\BIBentryALTinterwordstretchfactor}{4}
\providecommand{\BIBentryALTinterwordspacing}{\spaceskip=\fontdimen2\font plus
\BIBentryALTinterwordstretchfactor\fontdimen3\font minus
  \fontdimen4\font\relax}
\providecommand{\BIBforeignlanguage}[2]{{%
\expandafter\ifx\csname l@#1\endcsname\relax
\typeout{** WARNING: IEEEtran.bst: No hyphenation pattern has been}%
\typeout{** loaded for the language `#1'. Using the pattern for}%
\typeout{** the default language instead.}%
\else
\language=\csname l@#1\endcsname
\fi
#2}}
\providecommand{\BIBdecl}{\relax}
\BIBdecl

\bibitem{choi2016millimeter}
J.~Choi, V.~Va, N.~Gonzalez-Prelcic, R.~Daniels, C.~R. Bhat, and R.~W. Heath,
  ``Millimeter-wave vehicular communication to support massive automotive
  sensing,'' \emph{IEEE Commun. Mag.}, vol.~54, no.~12, pp. 160--167, 2016.

\bibitem{heath2016overview}
R.~W. Heath, N.~Gonzalez-Prelcic, S.~Rangan, W.~Roh, and A.~M. Sayeed, ``An
  overview of signal processing techniques for millimeter wave {MIMO}
  systems,'' \emph{IEEE J. Sel. Topics Signal. Process.}, vol.~10, no.~3, pp.
  436--453, 2016.

\bibitem{rappaport2013millimeter}
T.~S. Rappaport, S.~Sun, R.~Mayzus, H.~Zhao, Y.~Azar, K.~Wang, G.~N. Wong,
  J.~K. Schulz, M.~Samimi, and F.~Gutierrez, ``Millimeter wave mobile
  communications for 5{G} cellular: It will work!'' \emph{IEEE Access}, vol.~1,
  pp. 335--349, 2013.

\bibitem{rappaport2014millimeter}
T.~S. Rappaport, R.~W. Heath~Jr, R.~C. Daniels, and J.~N. Murdock,
  \emph{Millimeter wave wireless communications}.\hskip 1em plus 0.5em minus
  0.4em\relax Pearson Education, 2014.

\bibitem{festag2015standards}
A.~Festag, ``Standards for vehicular communication from {IEEE} 802.11 p to
  5{G},'' \emph{e \& i Elektrotechnik und Informationstechnik}, vol. 132,
  no.~7, pp. 409--416, 2015.

\bibitem{shah20185g}
S.~A.~A. Shah, E.~Ahmed, M.~Imran, and S.~Zeadally, ``5{G} for vehicular
  communications,'' \emph{IEEE Commun. Mag.}, vol.~56, no.~1, pp. 111--117,
  2018.

\bibitem{marzi2016compressive}
Z.~Marzi, D.~Ramasamy, and U.~Madhow, ``Compressive channel estimation and
  tracking for large arrays in mm-wave picocells,'' \emph{IEEE J. Sel. Topics
  Signal Process.}, vol.~10, no.~3, pp. 514--527, 2016.

\bibitem{ali2017millimeter}
A.~Ali, N.~Gonz{\'a}lez-Prelcic, and R.~W. Heath, ``Millimeter wave
  beam-selection using out-of-band spatial information,'' \emph{IEEE Trans.
  Wireless Commun.}, vol.~17, no.~2, pp. 1038--1052, 2017.

\bibitem{santa2009sharing}
J.~Santa and A.~F. Gomez-Skarmeta, ``Sharing context-aware road and safety
  information,'' \emph{IEEE Pervasive Comput.}, vol.~8, no.~3, pp. 58--65,
  2009.

\bibitem{papadimitratos2009vehicular}
P.~Papadimitratos, A.~L. Fortelle, K.~Evenssen, R.~Brignolo, and S.~Cosenza,
  ``Vehicular communication systems: Enabling technologies, applications, and
  future outlook on intelligent transportation,'' \emph{IEEE Commun. Mag.},
  vol.~47, no.~11, pp. 84--95, 2009.

\bibitem{va2017inverse}
V.~Va, J.~Choi, T.~Shimizu, G.~Bansal, and R.~W. Heath, ``Inverse multipath
  fingerprinting for millimeter wave {V2I} beam alignment,'' \emph{IEEE Trans.
  Veh. Technol.}, vol.~67, no.~5, pp. 4042--4058, 2017.

\bibitem{wang:2019access}
Y.~Wang, A.~Klautau, M.~Ribero, A.~C. Soong, and R.~W. Heath, ``Mm{W}ave
  vehicular beam training with situational awareness using machine learning,''
  \emph{IEEE Access}, 2019.

\bibitem{gonzalez2017millimeter}
N.~Gonz{\'a}lez-Prelcic, A.~Ali, V.~Va, and R.~W. Heath, ``Millimeter-wave
  communication with out-of-band information,'' \emph{IEEE Commun. Mag.},
  vol.~55, no.~12, pp. 140--146, 2017.

\bibitem{myers2020deep}
N.~J. Myers, Y.~Wang, N.~Gonz{\'a}lez-Prelcic, and R.~W. Heath, ``Deep
  learning-based beam alignment in mmwave vehicular networks,'' in \emph{Proc.
  Int. Conf. Acoustics, Speech Signal Process. (ICASSP)}.\hskip 1em plus 0.5em
  minus 0.4em\relax IEEE, 2020, pp. 8569--8573.

\bibitem{ali2019passive}
A.~Ali, N.~Gonz{\'a}lez-Prelcic, and A.~Ghosh, ``Passive radar at the roadside
  unit to configure millimeter wave vehicle-to-infrastructure links,''
  \emph{arXiv preprint arXiv:1910.10817}, 2019.

\bibitem{ali2019millimeter}
------, ``Millimeter wave {V2I} beam-training using base-station mounted
  radar,'' in \emph{Proc. Radar Conf.}\hskip 1em plus 0.5em minus 0.4em\relax
  IEEE, 2019, pp. 1--5.

\bibitem{gonzalez2016radar}
N.~Gonz{\'a}lez-Prelcic, R.~M{\'e}ndez-Rial, and R.~W. Heath, ``Radar aided
  beam alignment in mm{W}ave {V2I} communications supporting antenna
  diversity,'' in \emph{Proc. Inf. Theory Appl. Workshop (ITA)}.\hskip 1em plus
  0.5em minus 0.4em\relax IEEE, 2016, pp. 1--7.

\bibitem{klautau2019lidar}
A.~Klautau, N.~Gonz{\'a}lez-Prelcic, and R.~W. Heath, ``{LIDAR} data for deep
  learning-based mm{W}ave beam-selection,'' \emph{IEEE Wireless Commun. Lett.},
  2019.

\bibitem{garcia2016location}
N.~Garcia, H.~Wymeersch, E.~G. Str{\"o}m, and D.~Slock, ``Location-aided
  mm-{w}ave channel estimation for vehicular communication,'' in \emph{Proc.
  Int. Workshop Signal Process Advances Wireless Commun. (SPAWC)}.\hskip 1em
  plus 0.5em minus 0.4em\relax IEEE, 2016, pp. 1--5.

\bibitem{kenney2011dedicated}
J.~B. Kenney, ``Dedicated short-range communications ({DSRC}) standards in the
  united states,'' \emph{Proc. IEEE}, vol.~99, no.~7, pp. 1162--1182, 2011.

\bibitem{chen2017vehicle}
S.~Chen, J.~Hu, Y.~Shi, Y.~Peng, J.~Fang, R.~Zhao, and L.~Zhao,
  ``Vehicle-to-everything ({V2X}) services supported by {LTE}-based systems and
  5{G},'' \emph{IEEE Commun. Standards Mag.}, vol.~1, no.~2, pp. 70--76, 2017.

\bibitem{wang2018mmwave}
Y.~Wang, K.~Venugopal, A.~F. Molisch, and R.~W. Heath, ``Mm{W}ave
  vehicle-to-infrastructure communication: Analysis of urban microcellular
  networks,'' \emph{IEEE Trans. Veh. Technol.}, vol.~67, no.~8, pp. 7086--7100,
  2018.

\bibitem{klautau20185g}
A.~Klautau, P.~Batista, N.~Gonz{\'a}lez-Prelcic, Y.~Wang, and R.~W. Heath,
  ``5{G} {MIMO} data for machine learning: Application to beam-selection using
  deep learning,'' in \emph{Proc. Inf. Theory Appl. Workshop (ITA)}.\hskip 1em
  plus 0.5em minus 0.4em\relax IEEE, 2018, pp. 1--9.

\bibitem{va2019online}
V.~Va, T.~Shimizu, G.~Bansal, and R.~W. Heath, ``Online learning for
  position-aided millimeter wave beam training,'' \emph{IEEE Access}, vol.~7,
  pp. 30\,507--30\,526, 2019.

\bibitem{mairal2010online}
J.~Mairal, F.~Bach, J.~Ponce, and G.~Sapiro, ``Online learning for matrix
  factorization and sparse coding,'' \emph{J. Machine Learning Research},
  vol.~11, no. Jan, pp. 19--60, 2010.

\bibitem{kivinen2004online}
J.~Kivinen, A.~J. Smola, and R.~C. Williamson, ``Online learning with
  kernels,'' \emph{IEEE Trans. Signal Process.}, vol.~52, no.~8, pp.
  2165--2176, 2004.

\bibitem{booth2019multi}
M.~B. Booth, V.~Suresh, N.~Michelusi, and D.~J. Love, ``Multi-armed bandit beam
  alignment and tracking for mobile millimeter wave communications,''
  \emph{IEEE Commun. Lett.}, vol.~23, no.~7, pp. 1244--1248, 2019.

\bibitem{sim2018online}
G.~H. Sim, S.~Klos, A.~Asadi, A.~Klein, and M.~Hollick, ``An online
  context-aware machine learning algorithm for 5{G} mm{W}ave vehicular
  communications,'' \emph{IEEE/ACM Trans. Network.}, vol.~26, no.~6, pp.
  2487--2500, 2018.

\bibitem{hashemi2018efficient}
M.~Hashemi, A.~Sabharwal, C.~E. Koksal, and N.~B. Shroff, ``Efficient beam
  alignment in millimeter wave systems using contextual bandits,'' in
  \emph{Proc. Conf. Comput. Commun.}\hskip 1em plus 0.5em minus 0.4em\relax
  IEEE, 2018, pp. 2393--2401.

\bibitem{gittins2011multi}
J.~Gittins, K.~Glazebrook, and R.~Weber, \emph{Multi-armed bandit allocation
  indices}.\hskip 1em plus 0.5em minus 0.4em\relax John Wiley \& Sons, 2011.

\bibitem{garivier2011upper}
A.~Garivier and E.~Moulines, ``On upper-confidence bound policies for switching
  bandit problems,'' in \emph{Int. Conf. Alg. Learning Theory}.\hskip 1em plus
  0.5em minus 0.4em\relax Springer, 2011, pp. 174--188.

\bibitem{ji2008bayesian}
S.~Ji, Y.~Xue, L.~Carin \emph{et~al.}, ``Bayesian compressive sensing,''
  \emph{IEEE Trans. Signal Process.}, vol.~56, no.~6, p. 2346, 2008.

\bibitem{imageprocess}
A.~C. Kak, M.~Slaney, and G.~Wang, ``Principles of computerized tomographic
  imaging,'' \emph{Medical Physics}, vol.~29, no.~1, pp. 107--107, 2002.

\bibitem{hassan2002new}
M.~Hassan-Ali and K.~Pahlavan, ``A new statistical model for site-specific
  indoor radio propagation prediction based on geometric optics and geometric
  probability,'' \emph{IEEE Trans. Wireless Commun.}, vol.~1, no.~1, pp.
  112--124, 2002.

\bibitem{seidel1994site}
S.~Y. Seidel and T.~S. Rappaport, ``Site-specific propagation prediction for
  wireless in-building personal communication system design,'' \emph{IEEE
  Trans. Veh. Technol.}, vol.~43, no.~4, pp. 879--891, 1994.

\bibitem{lim2017map}
Y.-G. Lim, Y.~J. Cho, Y.~Kim, and C.-B. Chae, ``Map-based millimeter-wave
  channel models: An overview, guidelines, and data,'' \emph{arXiv preprint
  arXiv:1711.09052}, 2017.

\bibitem{ray_tracing}
``Wireless insite,''
  {\url{https://www.remcom.com/wireless-insite-em-propagation-software/}}.

\bibitem{myers2019falp}
N.~J. Myers, A.~Mezghani, and R.~W. Heath~Jr, ``{FALP}: Fast beam alignment in
  mm{W}ave systems with low-resolution phase shifters,'' \emph{arXiv preprint
  arXiv:1902.05714}, 2019.

\bibitem{brady2013beamspace}
J.~Brady, N.~Behdad, and A.~M. Sayeed, ``Beamspace {MIMO} for millimeter-wave
  communications: System architecture, modeling, analysis, and measurements,''
  \emph{IEEE Trans. Antennas Propag.}, vol.~61, no.~7, pp. 3814--3827, 2013.

\bibitem{kak2002principles}
A.~C. Kak, M.~Slaney, and G.~Wang, ``Principles of computerized tomographic
  imaging,'' \emph{Medical Physics}, vol.~29, no.~1, pp. 107--107, 2002.

\bibitem{lancaster2005chi}
H.~O. Lancaster and E.~Seneta, ``Chi-square distribution,'' \emph{Encyclopedia
  of Biostatistics}, vol.~2, 2005.

\bibitem{diamond2016cvxpy}
S.~Diamond and S.~Boyd, ``{CVXPY}: A {P}ython-embedded modeling language for
  convex optimization,'' \emph{The J. Mach. Learning Research}, vol.~17, no.~1,
  pp. 2909--2913, 2016.

\bibitem{gerchberg1972practical}
R.~W. Gerchberg, ``A practical algorithm for the determination of phase from
  image and diffraction plane pictures,'' \emph{Optik}, vol.~35, pp. 237--246,
  1972.

\bibitem{auer2002finite}
P.~Auer, N.~Cesa-Bianchi, and P.~Fischer, ``Finite-time analysis of the
  multi-armed bandit problem,'' \emph{Machine learning}, vol.~47, no. 2-3, pp.
  235--256, 2002.

\bibitem{book:banditalg}
T.~Lattimore and C.~Szepes{\'v}ari, \emph{Bandit Algorithms}.\hskip 1em plus
  0.5em minus 0.4em\relax Reading, MA: Addison-Wesley, 1972.

\bibitem{tropp2007signal}
J.~A. Tropp and A.~C. Gilbert, ``Signal recovery from random measurements via
  orthogonal matching pursuit,'' \emph{IEEE Trans. Inf. Theory}, vol.~53,
  no.~12, pp. 4655--4666, 2007.

\bibitem{mahalanobis1936generalized}
P.~C. Mahalanobis, ``On the generalized distance in statistics.''\hskip 1em
  plus 0.5em minus 0.4em\relax National Institute of Science of India, 1936.

\bibitem{beran1977minimum}
R.~Beran \emph{et~al.}, ``Minimum hellinger distance estimates for parametric
  models,'' \emph{The annals of Statistics}, vol.~5, no.~3, pp. 445--463, 1977.

\end{thebibliography}
